\documentclass[11pt,preprint,flushrt]{aastex}
\usepackage{cranm_timesv}
\usepackage{multicol}
\usepackage{epsfig}
\slugcomment{{\em Astrophys.\  J.} (2003), in press}

\newcommand{\para}{\parallel}
\shorttitle{Alfv\'{e}nic Turbulence in the Solar Corona}
\shortauthors{Cranmer and van Ballegooijen}

\begin{document}

\title{Alfv\'{e}nic Turbulence in the Extended Solar Corona:
Kinetic Effects and Proton Heating}

\author{S. R. Cranmer and A. A. van Ballegooijen}
\affil{Harvard-Smithsonian Center for Astrophysics,
60 Garden Street, Cambridge, MA 02138}

\begin{abstract}
\small
\baselineskip=12.7pt
We present a model of magnetohydrodynamic (MHD) turbulence
in the extended solar corona that contains the effects of
collisionless dissipation and anisotropic particle heating.
Recent observations have shown that preferential heating
and acceleration of positive ions occurs in the first
few solar radii of the high-speed solar wind. 
Measurements made by the Ultraviolet Coronagraph Spectrometer
aboard {\em SOHO} have revived interest in the idea that ions 
are energized by the dissipation of ion cyclotron resonant
waves, but such high-frequency (i.e., small wavelength)
fluctuations have not been observed.
A turbulent cascade is one possible way of generating 
small-scale fluctuations from a pre-existing population of
low-frequency MHD waves. 
We model this cascade as a combination of advection and
diffusion in wavenumber space.
The dominant spectral transfer occurs in the direction
perpendicular to the background magnetic field.
As expected from earlier models, this leads to a highly
anisotropic fluctuation spectrum with a rapidly decaying
tail in the parallel wavenumber direction.
The wave power that decays to high enough frequencies to
become ion cyclotron resonant depends on the relative 
strengths of advection and diffusion in the cascade.
For the most realistic values of these parameters, though,
there is insufficient power to heat protons and heavy ions.
The dominant oblique fluctuations (with dispersion properties
of kinetic Alfv\'{e}n waves) undergo Landau damping, which
implies strong parallel electron heating.
We discuss the probable nonlinear evolution of the 
electron velocity distributions into parallel beams and 
discrete phase-space holes (similar to those seen in 
the terrestrial magnetosphere) which can possibly heat 
protons via stochastic interactions.    
\end{abstract}


\section{Introduction}

\setlength{\columnsep}{0.67cm}
\begin{multicols}{2}

\small
\baselineskip=12.7pt
\parindent=0.17in
In order to produce the hot ($10^6$~K) solar corona, a fraction
of the mechanical energy in the Sun's internal convective motions
must be converted into heat above the photosphere.
Despite more than a half century of investigation, though,
the precise physical processes that lead to coronal heating
and the subsequent acceleration of the solar wind are not known.
At the coronal base, different combinations of mechanisms
(e.g., magnetic reconnection, turbulence, wave dissipation, and
plasma instabilities) are probably responsible for the varied
appearance of coronal holes, quiet regions, loops, and X-ray
bright points (Priest et al.\  2000).
In the open magnetic flux tubes that feed the solar wind,
additional heating at heliocentric distances greater than
about 2 solar radii ($R_{\odot}$) is believed to be needed
(e.g., Leer, Holzer, \& {Fl\aa} 1982; Parker 1991).
This paper investigates the consequences of a promising
mechanism for extended coronal heating: the kinetic dissipation
of driven magnetohydrodynamic (MHD) turbulence in the strong
background magnetic field of the accelerating solar wind.

The necessity for gradual energy deposition at large distances
from the coronal base comes from three sets of empirical
constraints (see also Cranmer 2002).
First, pressure-driven models of the high-speed component
($v \gtrsim 750$ km~s$^{-1}$) of the solar wind cannot be
made consistent with the relatively low inferred electron
temperatures in coronal holes
($T_{e} < 1.5 \times 10^6$ K) without additional
energy or momentum deposition.
Second, spacecraft in the interplanetary medium have measured
radial gradients in proton and electron temperatures that
are substantially shallower than predicted from pure
adiabatic expansion (Phillips et al.\  1995;
Richardson et al.\  1995).
Similar measurements of radial growth of the proton
magnetic moment between the orbits of Mercury and the
Earth (Schwartz \& Marsch 1983; Marsch 1991) point to
specific forms of gradual energy addition.
Third, the Ultraviolet Coronagraph Spectrometer (UVCS) aboard
the {\em Solar and Heliospheric Observatory} ({\em SOHO})
measured extremely high heavy ion temperatures, faster
bulk ion outflow compared to protons, and strong
anisotropies (with $T_{\perp} \gg T_{\para}$) in ion
velocity distributions in the extended corona
(Kohl et al.\  1997, 1998, 1999; Noci et al.\  1997;
Cranmer et al.\  1999b; Giordano et al.\  2000).

The list of possible physical processes responsible for
extended coronal heating is limited severely by the fact that
Coulomb collisions are extremely weak above 2 to 3 $R_{\odot}$
and that ions receive more energy than electrons (with
$T_{\rm ion} \gg T_{p} \gtrsim T_{e}$).
Also, most suggested mechanisms involve the transfer of
energy from propagating fluctuations---such as waves,
shocks, or turbulent eddies---to the particles.
This broad consensus has arisen because the ultimate
source of energy must be solar in origin, and thus it
must somehow be transmitted out to the distances where
the heating occurs (see, e.g., Hollweg 1978;
Tu \& Marsch 1995).
The most common wave damping mechanisms proposed for the
coronal base (resistivity, viscosity, and thermal
conductivity) seem to be ruled out in the extended corona
because of the relative unimportance of collisions.

Wave-particle interactions are natural alternatives to
collisional processes and have been studied in a solar wind
context for several decades (Barnes 1968; Toichi 1971;
Abraham-Shrauner \& Feldman 1977; Hollweg \& Turner 1978;
Marsch, Goertz, \& Richter 1982; Isenberg \& Hollweg 1983;
Hollweg 1986; Tu 1987, 1988; Axford \& McKenzie 1992).
The {\em SOHO} observations discussed above have given rise
to a resurgence of interest in the collisionless dissipation
of ion cyclotron waves as a potentially important mechanism
in the acceleration region of the high-speed wind
(e.g., McKenzie et al.\  1995; Tu \& Marsch 1997, 2001;
Hollweg 1999b; Li et al.\  1999;
Galinsky \& Shevchenko 2000; Cranmer 2000, 2001, 2002;
Hollweg \& Isenberg 2002; Vocks \& Marsch 2002).
In the extended corona, the cyclotron or Larmor frequencies
of positive ions range from 10 to 10$^4$ Hz, but the
oscillation frequencies observed on the solar surface
(dominated by convection) are typically $\sim$0.01 Hz.
Thus, any wave generation mechanism seems to require bridging
a gap of many orders of magnitude in frequency or wavenumber.

Axford \& McKenzie (1992) suggested that high-frequency
Alfv\'{e}n waves could be generated during small-scale
reconnection events in the rapidly evolving supergranular
network.
These waves would propagate up through the corona until they
reach the cyclotron resonance radii of various ions, then
damp over a very short distance (see also Schwartz, Feldman,
\& Gary 1981).
Tu \& Marsch (1997) and Marsch \& Tu (1997) presented solar
wind models based on the idea that a launched power spectrum
becomes ``swept up'' and eroded by the slow radial decrease
of the cyclotron frequency.
The general scenario of cyclotron sweeping has been called into
question by Cranmer (2000), Hollweg (2000), and Leamon et
al.\  (2000), but its relative importance in relation to other
physical processes has not yet been fully established (see also
Tu \& Marsch 2001).

Alternatively, there have been numerous ``local'' wave
generation scenarios proposed that involve the transfer of
energy from other sources into the ion cyclotron mode at a
range of heights in the acceleration region of the solar wind.
Into this class of models fall the processes of MHD turbulent
cascade, kinetic plasma instabilities (driven by non-Maxwellian
velocity distributions or spatial gradients), or wave mode
conversion driven by reflection or refraction (e.g.,
Hollweg 1986; Matthaeus et al.\  1999;
Kaghashvili \& Esser 2000; Markovskii 2001; Moran 2002).
Most of these local wave generation models rely on the Sun
launching a sufficiently intense spectrum of low-frequency
(i.e., with periods longer than $\sim$5 min) Alfv\'{e}n or
fast-mode MHD waves that do not damp within a solar radius
of the surface.
At present, however, there are few empirical constraints on
the generation mechanisms, exact propagation modes, or power
levels of ion cyclotron waves in the corona.

In this paper we study the consequences of anisotropic
MHD turbulent cascade as a possible generation mechanism
for ion cyclotron waves in the extended corona.
We also examine the turbulent cascade as a source of other
forms of energy---such as parallel electron beams and
phase-space holes---that could also lead to ion heating
in the corona.
We model the cascade as a diffusion process in
three-dimensional wavenumber space and recover the basic
form of the Goldreich \& Sridhar (1995) spectral anisotropy.
This solution for the turbulent power spectrum is coupled
to a general Alfv\'{e}n wave dispersion relation in order to
compute the relative amounts of heating given to protons and
electrons.
The dispersive properties of the highly oblique kinetic
Alfv\'{e}n wave (KAW; see St\'{e}fant 1970; Hasegawa 1976;
Lysak \& Lotko 1996; Hollweg 1999a; Stasiewicz et al.\  2000)
turn out to be key drivers of the plasma conditions of the
extended corona. 
The KAW has been studied in a coronal and solar wind context by
Dobrowolny \& Torricelli-Ciamponi (1985), Song (1996),
Leamon et al.\  (1998, 1999, 2000), Shukla et al.\  (1999),
and Voitenko \& Goossens (2002).

The remainder of this paper is organized as follows.
In {\S}~2 we develop a semianalytical model of anisotropic
MHD turbulence by extending the wavenumber diffusion picture
of Zhou \& Matthaeus (1990) to three dimensions.
In {\S}~3, proton and electron heating rates that are
consistent with the derived power spectrum are computed
from a Vlasov-Maxwell kinetic dispersion relation.
For the realistic case that KAWs heat electrons in the
direction parallel to the magnetic field, {\S}~4 follows
a speculative chain of reasoning that leads to the generation
of electron phase-space holes (EPHs) that can heat protons
and heavy ions via Coulomb-like collisions.
Conclusions, remaining questions, and implications for
future spectroscopic observations are summarized in {\S}~5.

\section{Turbulent Cascade in the Extended Corona}

In this section we derive the Fourier power spectrum of
turbulent Alfv\'{e}nic fluctuations as a function of the
wavevector ${\bf k}$ in the low plasma-beta extended corona.
We assume that the turbulence is driven at large scales
by waves that originate at the Sun, and that some process
(such as reflection) generates inward-propagating waves to
excite a cascade (Matthaeus et al.\  1999).
The precise generation mechanisms of the low-frequency waves
at the solar surface are unimportant in the context of this
paper (see, e.g., Roberts 2000).
The outward-propagating waves enter the corona and transform,
in part, to a mixed population of outward and inward modes.
Only when inward and outward wave packets are allowed to
``collide'' can nonlinear couplings to higher wavenumber
occur.
The resulting cascade is a time-steady transfer of energy from
large to small spatial scales that is believed to progress
by successive magnetic reconnection in the presence of the
strong coronal ``guide field'' (e.g.,
Matthaeus \& Lamkin 1986; Lazarian \& Vishniac 1999).
We assume that the turbulence becomes fully developed
on time scales that are short compared to the bulk solar wind
outflow and the geometrical expansion of open flux tubes.
This allows us to model the turbulence as spatially homogeneous
in a small volume element (with constant density and magnetic
field strength) in the extended corona.

\subsection{Definitions of Fluctuation Quantities}

The energy density of fluctuations in a magnetized plasma is
described most generally as a sum of electric, magnetic, and
kinetic components.
For linear (i.e., small amplitude) fluctuations, the
total fluctuation energy density $\delta U$ is given by
\begin{displaymath}
  \delta U \,\,\, = \,\,\,
  \frac{\langle \delta {\bf E}^{2} \rangle}{8\pi} \, + \,
  \frac{\langle \delta {\bf B}^{2} \rangle}{8\pi} \, - \,
\end{displaymath}
\begin{equation}
  \sum_{s} \int \! d^{3} {\bf v} \, \left( \frac{1}{2} m_{s}
  | {\bf v}^{2} | \right) \, \frac{\langle \delta f^2 \rangle_s}
  {{\bf v} \cdot \partial f_{0} / \partial {\bf v}}
  \label{eq:dU}
\end{equation}
where angle brackets denote averages over times much longer
than the fluctuation time scales, $\delta {\bf E}$ is the
perturbed electric field, and $\delta {\bf B}$ is the perturbed
magnetic field (e.g., Bernstein 1958).
The velocity distribution function of particle species $s$
is assumed to be the sum of an undisturbed, stationary
(zero-order) state $f_0$ and a small, linear (first-order)
perturbation $\delta f$.
The masses and velocities of the particles are denoted $m_s$
and ${\bf v}$, and the integration above is taken over all
particle velocities for each species.
For simplicity, we assume a fully ionized plasma
composed only of protons and electrons (see {\S}~5 for a
discussion of the impact of heavy ions).
The quantity $\delta U$ above is constructed as the simplest
second-order and positive-definite quantity that is conserved
exactly as a consequence of the fully nonlinear Vlasov
equation (Davidson 1983).

If the zero-order proton and electron velocity distributions
are assumed to be Maxwellian, the kinetic energy density term
in eq.~(\ref{eq:dU}) can be expressed as the sum of fluctuations
in bulk velocity and density (often referred to as ``kinetic''
and ``thermal'' energies, respectively).
We assume that these two contributions to the total kinetic
energy are the dominant ones, even when $f_0$ departs from
a Maxwellian form.
Thus, the total energy density is given by
\begin{displaymath}
  \delta U \,\,\, = \,\,\,
  \frac{\langle \delta {\bf E}^{2} \rangle}{8\pi} \, + \,
  \frac{\langle \delta {\bf B}^{2} \rangle}{8\pi} \, + \,
\end{displaymath}
\begin{equation}
  \sum_{s} \left( \frac{1}{2} m_{s} n_{s}
  \langle \delta {\bf v}^{2} \rangle_{s} \, + \,
  \frac{1}{2} k_{\rm B} T_{s} 
  \frac{\langle \delta n^{2} \rangle_s}{n_s} \right)
  \,\, ,
  \label{eq:dUmax}
\end{equation}
where $n_s$ and $T_s$ are the zero-order number densities and
temperatures, $\delta {\bf v}$ and $\delta n$ are the
first-order velocity and number density perturbations,
and $k_{\rm B}$ is Boltzmann's constant.
For a nonrelativistic plasma, the electric field term
is usually negligible in comparison with the other terms.
Plasmas with sufficiently strong magnetic fields tend to
be dominated by incompressible magnetohydrodynamic
(MHD) fluctuations (see, e.g.,
Goldstein, Roberts, \& Matthaeus 1995),
which exhibit energy density equipartition between their
magnetic and proton velocity fluctuations,
\begin{equation}
  \frac{\delta U_{\rm MHD}}{\rho} \, \approx \, \frac{1}{2}
  \left( \frac{\langle \delta {\bf B}^{2} \rangle}{4\pi\rho}
  \, + \, \langle \delta {\bf v}^{2} \rangle_{p} \right)
  \,\, \approx \,\,
  \langle \delta {\bf v}^{2} \rangle_{p}
  \,\,\, ,
\end{equation}
where $\rho$ is the total mass density, approximately equal to
its proton contribution $m_{p} n_{p}$.
At the base of the solar corona, the total velocity amplitude
$(\delta U / \rho)^{1/2}$ is believed to be of order
20 to 40 km s$^{-1}$ (e.g., Mariska, Feldman, \& Doschek 1978;
Chae, Sch\"{u}hle, \& Lemaire 1998).
This population of waves is expected to be mostly propagating
upwards from the Sun, with a radial amplitude dependence
close to that predicted by the conservation of wave action
(Hollweg 1973; Jacques 1977; Banerjee et al.\  1998;
Esser et al.\  1999).

In this paper, we need to evaluate several of the terms in
the total energy density $\delta U$ independently.
However, the turbulence as a whole is describable in terms
of the total fluctuation power spectrum $W$, defined
as the energy density per unit volume in
three-dimensional wavenumber space and scaled as follows
\begin{equation}
  \frac{\delta U}{\rho} \, = \,
  \int d^{3} {\bf k} \,\,
  W (k_{\para}, k_{\perp}, t) \,\,\, .
  \label{eq:Wdef}
\end{equation}
MHD turbulence is dominated by a cascade of fluctuation energy
in the direction {\em perpendicular} to the background
magnetic field, so let us also define the reduced power
spectrum quantity $W_{\perp}$,
\begin{equation}
  W_{\perp} (k_{\perp}, t) \, \equiv \,
    k_{\perp}^{2} \int_{-\infty}^{+\infty} dk_{\para} \,\,
    W (k_{\para}, k_{\perp}, t)
    \,\,\, ,
\end{equation}
which is proportional to the total energy density per unit
$\ln k_{\perp}$, integrated over the parallel wavenumber
$k_{\para}$.
The quantity $W_{\perp}$ has units of velocity squared, and
is related to the total energy density via
\begin{equation}
  \delta U \, = \, 2\pi \rho
  \int_{0}^{+\infty} \frac{dk_{\perp}}{k_{\perp}} \,
  W_{\perp} (k_{\perp})
  \,\,\, .
  \label{eq:Wperpint}
\end{equation}
Note that $k_{\para}$ can be both positive (for outward
propagating waves) and negative (for inward propagating
waves), but the perpendicular wavenumber $k_{\perp}$ is
positive-definite.

The energy density spectra defined above are appropriate for
following the flow of the {\em total} energy in wavenumber
space, but they do not necessarily track the dominant turbulent
motions on all spatial scales.
One important example is the case of large-$k_{\perp}$
kinetic Alfv\'{e}n waves (KAW) that have wavelengths
larger than the mean electron Larmor radius but smaller
than the mean proton Larmor radius.
These fluctuations are ``felt'' by all electrons in a similar
manner as longer-wavelength MHD waves.
However, some protons execute Larmor gyrations on a spatial
scale larger than that of the fluctuations, and thus they
become uncoupled from the waves.
(Only the slowest protons---in the core of their velocity
distribution---have small enough Larmor orbits to
remain coupled to the electrons and waves.)
Thus, even though the massive protons carry the bulk of
the kinetic energy, it is the electron term
$\langle \delta {\bf v}^{2} \rangle_{e}$ that
most accurately describes the smallest-scale fluctuations.
Let us then define $v_{\perp}^{2}$ as the power spectrum of
electron kinetic energy fluctuations in the perpendicular
direction, which is defined in a similar manner as
eq.~(\ref{eq:Wperpint}),
\begin{equation}
  \langle \delta {\bf v}_{\perp}^{2} \rangle_{e} \, = \, 2\pi
  \int_{0}^{+\infty} \frac{dk_{\perp}}{k_{\perp}} \,
  v_{\perp}^{2} (k_{\perp})
\end{equation}
where only the perpendicular components of the electron
velocity fluctuation are considered.%
\footnote{For kinetic Alfv\'{e}n waves, the parallel electron
velocity fluctuation begins to dominate when $k_{\perp}$
becomes of the same order as the inverse ion inertial length
(see Hollweg 1999a). However, we restrict the definition of
$v_{\perp}$ to the perpendicular fluctuations in order to
be consistent with anisotropic MHD turbulence theory.}
In the low-$k_{\perp}$ MHD limit, where protons and electrons
oscillate with the same velocity, $v_{\perp}^{2}$ is equal
to $W_{\perp}$.  In general, we define
\begin{equation}
  v_{\perp}^{2} (k_{\perp}) \, = \,
  \phi_{e} (k_{\perp}) \, W_{\perp} (k_{\perp})
  \label{eq:phidef}
\end{equation}
where $\phi_e$ describes the departure from ideal MHD
energy equipartition, and is equal to 1 in the MHD limit.
Formally, $\phi_e$ is the ratio of kinetic energy density in
perpendicular electron motions to the total energy density
of the fluctuations at a specified wavenumber.
The inverse quantity $\phi_{e}^{-1}$ is also approximately
equal to the number fraction of protons that remain coupled
to the fluctuations at the specified wavenumber.
In {\S}~3.1 we derive the exact behavior of $\phi_e$ as
a function of $k_{\perp}$, but for simplicity we
present the following approximation,
\begin{equation}
  \phi_{e} \, \approx \, 1 + k_{\perp}^{2} R_{p}^2
  \,\,\, ,
  \label{eq:fedef}
\end{equation}
where $R_p$ is the mean proton gyroradius, defined as the
ratio of the proton most-probable speed
$w_{p} = (2 k_{\rm B} T_{p} / m_{p})^{1/2}$ to the proton
cyclotron frequency $\Omega_{p} = e B / m_{p} c$.
The above approximate form for $\phi_e$ was motivated by a
simple estimate of the fraction of protons that remain
coupled to $k_{\perp}$-scale fluctuations; i.e.,
\begin{equation}
  \frac{1}{\phi_e} \approx \left.
  \left[ \int_{0}^{V_{\rm max}} d^{2} v_{\perp} \,
  f_{p} (v_{\perp}) \right]
  \right/
  \left[ \int_{0}^{\infty} \! d^{2} v_{\perp} \,
  f_{p} (v_{\perp}) \right]
  \,\, .
\end{equation}
For a Maxwellian velocity distribution with an effective
``core'' defined by
$v_{\perp} < V_{\rm max} \equiv \Omega_{p} / k_{\perp}$,
the above expression yields a fraction of 1 in the limit
$k_{\perp} R_{p} \ll 1$ and a fraction of
$(k_{\perp} R_{p})^{-2}$ in the limit
$k_{\perp} R_{p} \gg 1$.
Eq.~(\ref{eq:fedef}) is a simple function that smoothly
bridges both limits.

\subsection{Dominant Two-Dimensional Cascade}

We describe the cascade of energy in MHD turbulence as a
combination of advection and diffusion in wavenumber space.
Chandrasekhar (1943) developed the general statistical theory
of representing stochastic processes---such as Brownian
``random walks''---as effectively diffusive.
Leith (1967) proposed a general advection-diffusion equation
for the evolution of fluctuation power in isotropic
hydrodynamic turbulence.
Leith's chief assumption was that the spectral cascade term
must only redistribute energy in wavenumber space and should
not alter its total magnitude.
Thus, the spectral cascade term was written as the divergence
of a flux-like quantity, where the primary contribution to
the flux is from nearby regions of wavenumber space (e.g.,
Kolmogorov 1941; Obukhov 1941).
Pao (1965), Eichler (1979), Tu, Pu, \& Wei (1984), and
Tu (1988) modeled the spectral transfer as pure advection
by deriving dimensionally consistent flux quantities that
are scalar functions of the local wave power and wavenumber.
Zhou \& Matthaeus (1990) applied Leith's (1967) cascade
phenomenology to MHD turbulence and derived diffusion
coefficients consistent with both the Kolmogorov (1941) and
Kraichnan (1965) energy transfer rates.

It has been known for several decades that MHD turbulence
in the presence of a background magnetic field develops
a strong wavenumber anisotropy.
Both numerical simulations and analytic descriptions such
as ``reduced MHD'' (RMHD) have indicated that cascade
from large to small spatial scales proceeds mainly in
the two-dimensional plane perpendicular to the background
field ${\bf B}_0$ (e.g., Strauss 1976; Montgomery 1982;
Shebalin, Matthaeus, \& Montgomery 1983;
Matthaeus et al.\  1996; Goldreich \& Sridhar 1995, 1997;
Maron \& Goldreich 2001; Bhattacharjee \& Ng 2001;
Cho, Lazarian, \& Vishniac 2002).
We apply the Zhou \& Matthaeus (1990) wavenumber diffusion
idea to this largely two-dimensional transfer of energy
from small $k_{\perp}$ to large $k_{\perp}$.

We model the interplay of wave injection (at small $k_{\perp}$),
cascade, and damping (at large $k_{\perp}$) with an
advection-diffusion equation similar to those used by
Zhou \& Matthaeus (1990), Miller, LaRosa, \& Moore (1996),
and Stawicki, Gary, \& Li (2001).
The $k_{\para}$-integrated fluctuation power is assumed to
obey the following transport equation,
\begin{displaymath}
  \frac{\partial W_{\perp}}{\partial t} \, = \,
  \frac{\partial}{\partial x} \left[
  \frac{1}{\tau_s} \left( -\beta W_{\perp} + \gamma
  \frac{\partial W_{\perp}}{\partial x} \right) \right]
\end{displaymath}
\begin{equation}
  \, + \, S (k_{\perp}) \, - \, D (k_{\perp}) W_{\perp}
  \,\,\, ,
  \label{eq:dWdt}
\end{equation}
where $x = \ln k_{\perp}$, $\tau_s$ is a characteristic
spectral transfer time, and $\beta$ and $\gamma$ are
dimensionless constants.
The $S$ term describes the source of energy at low
$k_{\perp}$, which presumably propagates up from the Sun
in the form of low-frequency Alfv\'{e}n waves.
The $D$ term describes the dissipation of energy
at high $k_{\perp}$.
The local spectral transfer time $\tau_s$ is assumed to be
equal to the nonlinear time scale $(k_{\perp} v_{\perp})^{-1}$
(Batchelor 1953; Goldreich \& Sridhar 1995).
We use the perpendicular electron spectrum
$v_{\perp} (k_{\perp})$ to track the turbulent motions on all
scales, which introduces a factor of $\phi_{e}^{1/2}$ into the
advection-diffusion equation when $\tau_s$ is written
in terms of $W_{\perp}$.

Note that we did not include in eq.~(\ref{eq:dWdt}) any
large-scale transport terms that depend on radial gradients
of the coronal plasma parameters.
This is in accord with the assumption discussed above
that the cascade acts on a time scale that is short
in comparison to representative flow times (due to
wind advection, wave propagation, or geometrical expansion)
over a coronal scale height.
The interaction between large-scale ``sweeping'' terms in
the corona and local turbulent cascade effects is discussed
in detail by Leamon et al.\  (2000).

The nonlinear cascade process modeled in eq.~(\ref{eq:dWdt})
reflects some of the physics of wave dispersion and damping,
but not all.
The spectral transfer time $\tau_s$ behaves differently in
the ideal MHD regime ($\phi_{e} \approx 1$) and in the KAW
regime ($\phi_{e} \propto k_{\perp}^{2}$).
Moreover, the dissipation term $D$ in these regimes behaves
as a power-law in $k_{\perp}$, thus producing a phenomenology
very similar to that of viscosity in fluid turbulence
(e.g., Pao 1965; Leith 1967).
However, we do not include high-frequency (proton cyclotron)
dispersive effects in the cascade term, even though we do
include them in the dissipation term.%
\footnote{One can avoid possible inconsistencies of this kind
by assuming the wavelength ranges of cascade and dissipation
are separated from one another.  However, this allows only
the computation of the total energy fluxes and not a detailed
comparison of relative proton/electron and parallel/perpendicular
heating rates; see also Leamon et al.\  (1999, 2000).}
Such effects in the cascade term are expected to be negligible
because of the very limited extent of the ``dispersion range''
in $k_{\para}$ (see the closely-spaced solid contours in
Figure 4a, below).
Therefore, cyclotron dispersion cannot substantially affect
the rate of energy transport from low to high $k_{\para}$.
In future work we intend to explore more self-consistent
cascade phenomenologies for plasmas of arbitrary $k_{\para}$
and $k_{\perp}$.

In order to specify the $\beta$ and $\gamma$ constants as
independent advection and diffusion strengths, we follow the
statistical analysis of van Ballegooijen (1986).
In the ``random walk'' limit that individual turbulent
displacements are small compared to the outer-scale
correlation length, van Ballegooijen (1986) found that
$\beta \approx \gamma$.
We assume for simplicity that $\beta$ and $\gamma$ are
constants (i.e., independent of wavenumber).
We also assume that $\beta$ and $\gamma$ are of order unity
because the primary physics of the cascade is encapsulated
in the spectral transfer time $\tau_s$.
Our baseline model thus assigns $\beta = \gamma = 1$, and
other models vary either $\beta$ or $\gamma$ while keeping
the other equal to 1.
The most appropriate values of these constants to use in
the extended solar corona should be determined ultimately
from the analysis of numerical simulations or laboratory
experiments.

Figure 1 shows representative time-steady solutions to
eq.~(\ref{eq:dWdt}) computed using a Crank-Nicholson
numerical finite differencing scheme.
The Appendix contains further details about the solution
method and the definition of supplementary quantities.
The low-wavenumber source term $S (k_{\perp})$ was
defined as a narrow Gaussian function that injects most
of its energy at an outer-scale length
$L_{\rm out} \sim k_{\rm out}^{-1}$ that we specified as
approximately 3800 km.
This value is consistent with the photospheric granulation
scale (1000 km) expanded superradially in a polar coronal
hole out to a distance of 2 $R_{\odot}$ (i.e.,
$L_{\rm out} \propto B_{0}^{-1/2}$; see Hollweg 1986).
The coronal hole magnetic field structure is assumed
to follow the dipole/quadrupole/current-sheet model of
Banaszkiewicz, Axford, \& McKenzie (1998).
Note also that a similar outer-scale length ($\sim$1200 km
at the base) was derived---independently of granulation
observations---by Chae et al.\  (1998), who balanced an
empirical turbulent heating rate by local radiative losses
in the transition region and low corona.
We assume that the wave energy at the outer scale is generated
at the solar surface, propagates outward into the corona, and
undergoes some degree of transformation into a mixed population
of outward and inward modes (by, e.g., reflection).
The dissipation term $D (k_{\perp})$ used in the models
shown in Figure 1 was assumed to be similar to that
of standard kinematic viscosity; i.e.,
$D \propto k_{\perp}^{2}$.
We plot the power spectra computed with three different
normalizing constants for the dissipation term.

The solutions for $W_{\perp} (k_{\perp})$ fall into
several regimes of wavenumber.
For $k_{\perp} \ll k_{\rm out}$, there is no significant
turbulent cascade, and the residual power decays like
$W_{\perp} \propto k_{\perp}^{\beta / \gamma}$ as a result
of the ever-present diffusion.
The power spectrum peaks around $k_{\rm out}$ at a maximum
value that depends on the integrated total energy.
In Figure 1, we specify
$(\delta U / \rho)^{1/2} = 50$ km s$^{-1}$, which results
in a maximum value of $W_{\perp}^{1/2}$ of about
10.2 km s$^{-1}$.
The factor-of-5 difference between these two quantities is
a simple consequence of the shape of $W_{\perp} (k_{\perp})$
around the peak at $k_{\rm out}$.
For $k_{\perp} > k_{\rm out}$, the power spectrum begins
to decrease in an MHD power-law inertial range, with
a Kolmogorov-Obukhov spectral index
$W_{\perp} \propto k_{\perp}^{-2/3}$
(the standard one-dimensional power spectrum is given
by $W_{\perp} / k_{\perp} \propto k_{\perp}^{-5/3}$; see
eq.~[\ref{eq:Wperpint}] and the detailed derivation in the
Appendix).
The spectral index in the inertial range does not depend
on $\beta$ or $\gamma$, but only on the wavenumber and
power dependence of $\tau_s$.
As $k_{\perp}$ approaches the inverse proton gyroradius,
however, the factor $\phi_e$ in $\tau_s$ becomes larger
than 1, and the inertial range spectral index steepens
from --2/3 to --4/3.
We refer to these wavenumber regions below as the MHD
and KAW inertial ranges, respectively
(see also discussion of the ``dispersion range'' in
interplanetary space by Stawicki et al.\  2001).
At high $k_{\perp}$, where $D$ becomes of the same order as
the local cascade rate $\tau_{s}^{-1}$, the power spectrum
dissipates rapidly.

\subsection{Extension into Parallel Wavenumber}

The previous section described the dynamics of turbulence as
a function of $k_{\perp}$ and ignored the behavior of the
full power spectrum $W$ as a function of $k_{\para}$.
The strong predicted anisotropy of MHD turbulence justifies
this approach to first order, but we are also concerned with
the possible ``leakage'' of power to high values of $k_{\para}$
and thus to high frequencies.
A substantial amount of power at the cyclotron frequencies
of positive ions would lead to significant preferential
ion heating.

Let us model the spectral transport in $k_{\para}$ with a
diffusion term similar to that assumed above.
It should be emphasized, however, that the actual parallel
transport is probably {\em nonlocal} in wavenumber space
(due to, e.g., intermittent high-order wave-wave interactions
and nonlinear steepening; see Medvedev 2000)
and thus should not be describable by local diffusion.
Our primary goal is to recover the phenomenology of the
anisotropic ``critical balance'' described by Goldreich
\& Sridhar (1995) and not to model the physics of nonlocal
turbulent transport in detail.
We thus propose the following advection-diffusion equation
for $W ({\bf k})$:
\begin{displaymath}
  \frac{\partial W}{\partial t} \, = \,
  \frac{1}{k_{\perp}} \frac{\partial}{\partial k_{\perp}}
  \left\{ k_{\perp} v_{\perp} \left[
  -\beta k_{\perp}^{2} W \, + \right. \right.
\end{displaymath}
\begin{equation}
  \left. \left. \gamma k_{\perp}
  \frac{\partial}{\partial k_{\perp}} \left( k_{\perp}^{2} W
  \right) \right] \right\} \, + \,
  \alpha k_{\perp} v_{\perp} \widetilde{k_{\para}}^{2}
  \left( \! \frac{\partial^{2} W}{\partial k_{\para}^2}
  \! \right) + S' - D W
  \label{eq:dEdt}
\end{equation}
which has been written in this way in order to recover
eq.~(\ref{eq:dWdt}) when each term above is multiplied by
$k_{\perp}^{2}$ and integrated over $k_{\para}$.
The dimensionless parameter $\alpha$, presumably of the
same order as $\beta$ and $\gamma$, determines the relative
strength of the effective parallel cascade.
The wavenumber scale is set by $\widetilde{k_{\para}}$,
which is a typical parallel wavenumber determined from the
condition of critical balance (i.e., strong coupling between
turbulent mixing motions and Alfv\'{e}n-wave motions along
the field).
Let us define the dimensionless ratio
\begin{equation}
  y \, \equiv \, \frac{\omega_r}{k_{\perp} v_{\perp}}
  \, = \, k_{\para} / \widetilde{k_{\para}}
  \,\,\,
  \label{eq:ydef}
\end{equation}
where $\omega_r$ is the real frequency of Alfv\'{e}n waves
as a function of $k_{\para}$ and $k_{\perp}$, and the
Goldreich-Sridhar condition of critical balance is
equivalent to the condition $y=1$.
The source term $S' ({\bf k})$ in eq.~(\ref{eq:dEdt}) is
related to the source term in eq.~(\ref{eq:dWdt}) as follows,
\begin{equation}
  S (k_{\perp}) \, = \,
  k_{\perp}^{2} \int_{-\infty}^{+\infty} dk_{\para} \,\,
  S' (k_{\para}, k_{\perp})  \,\,\, ,
\end{equation}
and we assume $D({\bf k})$ is dominated by the same kinetic
sources of damping that were assumed in {\S}~2.2.

The advection-diffusion equation presented above does not
depend on the sign of $k_{\para}$, and thus it produces 
diffusion for waves propagating in both directions along
the magnetic field.
In this paper we assume that $W$ is symmetric about
$k_{\para} = 0$  (i.e., that the net cross
helicity of the fluctuations is zero).
This is consistent with the analysis of Goldreich \&
Sridhar (1995), but is manifestly not true for the
lowest-$k_{\perp}$ (energy-containing) modes at the
outer scale in the corona.
For these large-scale waves, there should be more power on
the outward-propagating side ($k_{\para} > 0$) than on the
inward-propagating side.
However, as the cascade proceeds to higher wavenumbers, the
power in outward and inward propagating modes should become
increasingly mixed (see, e.g., Roberts et al.\  1992;
Vasquez, Markovskii, \& Hollweg 2002).
The main reason we are modeling $W({\bf k})$ in this paper
is to compute the kinetic dissipation and particle heating
{\em at high wavenumber,} so the assumption that the outward
and inward power is equal seems justifiable.

In order to evaluate the quantity $y$ defined above, we
estimate the dispersion relation for kinetic Alfv\'{e}n waves
(KAW) as $\omega_{r} \approx k_{\para} V_{A} \phi_{e}^{1/2}$.
This relation assumes frequencies much lower than the proton
cyclotron frequency (see {\S}~3.1 for a more complete
discussion of dispersion).
The multiplicative factor of $\phi_{e}^{1/2}$ represents
the lower inertia in the KAW regime due to the
decoupling of protons from the wave motions.
The coupled-proton fraction $1/\phi_e$ can be considered
to be proportional to the effective ``mass'' $m$ of a body
undergoing simple harmonic motion; the frequency of
oscillations for such a body scales as
$\omega^{2} \sim \kappa / m$, where $\kappa$ is the
effective ``spring constant'' determined here by the
Alfv\'{e}n speed $V_A$.
Because both the numerator and denominator in
eq.~(\ref{eq:ydef}) contain factors of $\phi_{e}^{1/2}$,
the quantity $y$ can be expressed as
\begin{equation}
  y \, \approx \, \frac{k_{\para} V_A}{k_{\perp} W_{\perp}^{1/2}}
  \,\,\, ,
\end{equation}
which is essentially the inverse of the Goldreich \& Sridhar
(1995) nonlinearity parameter $\zeta$.
We use this approximation to derive an analytic solution
for $W (k_{\para}, k_{\perp})$, but in {\S}~3.2 we replace
the above relation with the complete definition for $y$
(eq.~[\ref{eq:ydef}]) and use a numerical solution for
$\omega_r$.

In the MHD inertial range, $W_{\perp}$ is proportional to
$k_{\perp}^{-2/3}$, and thus
$y \propto k_{\para} k_{\perp}^{-2/3}$.
Turbulent eddies that conform to the condition of critical
balance ($y=1$) exhibit a wavenumber scaling of
$k_{\para} \propto k_{\perp}^{2/3}$, which implies they
become increasingly elongated in the direction of the magnetic
field as the cascade proceeds to higher wavenumber.
This spectral anisotropy is discussed in more detail by, e.g.,
Goldreich \& Sridhar (1995), Maron \& Goldreich (2001), and
Cho et al.\  (2002), and possible observational evidence of
such elongated inhomogeneities was presented by
Grall et al.\  (1997).
In the KAW inertial range, the elongation should grow slightly
stronger, with $y \propto k_{\para} k_{\perp}^{-1/3}$.
However, dissipation is believed to set in around
$k_{\perp} R_{p} \sim 1$, before too much of the KAW
inertial range can develop.

To solve eq.~(\ref{eq:dEdt}) we make the important assumption
that $W$ is a separable function of two variables:
$k_{\perp}$ and $y$.
Following Goldreich \& Sridhar (1995), we write the wave
power spectrum as
\begin{equation}
  W ( k_{\perp}, y ) \, = \,
  \frac{V_{A} W_{\perp}^{1/2}}{k_{\perp}^3} \, g(y)
  \label{eq:Wgdef}
\end{equation}
which was defined in this way to ensure that the dimensionless
function $g(y)$ is normalized to unity,
\begin{equation}
  \int_{-\infty}^{+\infty} dy \,\, g(y) \, = \, 1
  \,\,\, ,
  \label{eq:gnorm}
\end{equation}
and also that eqs.~(\ref{eq:Wdef})--(\ref{eq:Wperpint}) are
satisfied.
The solution for $W_{\perp} ( k_{\perp} )$ is discussed above
in {\S}~2.2, and the Appendix shows how a known solution for
$W_{\perp}$ allows eq.~(\ref{eq:dEdt}) to be approximated as
an ordinary differential equation for $g(y)$.
The solution to this differential equation in the inertial
range is
\begin{equation}
  g(y) \, \propto \, \left[ 1 +
  \frac{\gamma ( 1 - \zeta )^{2} y^{2}}{\alpha} \right]^{-n}
  \label{eq:kapfun}
\end{equation}
with the normalization determined by eq.~(\ref{eq:gnorm}), and
\begin{equation}
  n \, = \, \frac{(\beta / \gamma) + \zeta + 1}
  {2 (1 - \zeta)}
  \label{eq:enndef}
\end{equation}
(see Appendix for details).
The dimensionless exponent $\zeta$ is defined by
\begin{equation}
  2 \zeta \, = \, - \frac{k_{\perp}}{W_{\perp}}
  \frac{\partial W_{\perp}}{\partial k_{\perp}}
  \,\,\, ,
  \label{eq:zetadef}
\end{equation}
and in the MHD and KAW inertial ranges, $\zeta = 1/3$
and $2/3$, respectively.
The above solution for $g(y)$ superficially resembles a
generalized Lorentzian, or kappa distribution (Olbert 1969),
which is Gaussian for small arguments and approaches a
power-law tail for large arguments.
Cho et al.\  (2002) analyzed numerical turbulence simulations
and found that $g(y)$ may be fit reasonably well by decaying
exponential or Castaing functions.
However, the simulations do not seem to have sufficient
dynamic range in the high-$k_{\para}$ modes to resolve
a possible power-law tail as in eq.~(\ref{eq:kapfun}).

The relative amount of wave dissipation that heats protons
and heavy ions depends sensitively on the power that
diffuses to high frequencies; i.e., $y \gg 1$.
In this limit, the above analytic solution in the inertial
range behaves as $g \propto y^{-2n}$, which (after
normalization) no longer depends on the parallel diffusion
coefficient $\alpha$.
Thus, even when $y \gg 1$, the dominant diffusion still
comes from the perpendicular $\beta$ and $\gamma$ terms.
The major contribution of the $\alpha$ term is to
``bootstrap'' the transport of power from low to high
$k_{\para}$; the perpendicular transport then takes over
to determine the residual $k_{\para}$-dependence of
the power spectrum.
A key parameter in the $y$-dependence of the power-law tail
is the ratio $\beta / \gamma$, which is the relative
strength of perpendicular advection over that of diffusion.
We extend our baseline assumption of the previous section
and assume $\alpha = \beta = \gamma = 1$, but for cases
when the ratio $\beta / \gamma$ is varied we keep the two
{\em diffusion} coefficients equal to one another
($\alpha = \gamma$).

\section{Wave Dispersion and Dissipation}

\subsection{Vlasov-Maxwell Kinetic Theory}

Once the total fluctuation power spectrum $W ({\bf k})$ is
known, it becomes possible to compute the relevant particle
heating rates in the extended corona.
Despite the fact that the strong, critically balanced
turbulence described above is not very ``wavelike''
(i.e., a coherent wave survives for only about one period
before nonlinear processes transfer its energy to
smaller scales), we follow the standard procedure of
using the linear wave dispersion relation to compute
frequencies and dissipation rates (see also Miller et
al.\  1996; Quataert 1998; Leamon et al.\  1999).
The amplitudes of the waves in the dissipation range are
extremely small (i.e.,
$\langle \delta B^{2} \rangle \ll B_{0}^2$), so the
assumption of linearity is probably adequate.
In at least one other turbulent astrophysical plasma---the
interstellar medium---a comprehensive study of various
dissipation mechanisms (Spangler 1991) showed that linear
damping processes are most likely to be stronger than
nonlinear damping processes.

We solve the linear dispersion relation for the real and
imaginary parts of the frequency in the solar-wind frame
($\omega = \omega_{r} + i \omega_{i}$) assuming a known
real wavevector ${\bf k}$.
The Vlasov-Maxwell dispersion relation,
\begin{equation}
  \frac{c^2}{\omega^2} \left[ {\bf k} \times \left(
  {\bf k} \times {\bf E} \right) \right] \, + \,
  \widetilde{\mbox{\boldmath $\epsilon$}} \cdot {\bf E}
  \, = \, 0  \,\,\, ,
\end{equation}
is derived by linearizing and Fourier transforming Maxwell's
equations (see, e.g., Stix 1992;
Krauss-Varban, Omidi, \& Quest 1994; Brambilla 1998).
Above, ${\bf E}$ is the perturbed electric field, $c$ is the
speed of light, and $\widetilde{\mbox{\boldmath $\epsilon$}}$
is the dielectric tensor constrained by the
linearized Vlasov equation.
The dispersion relation is essentially a $3 \times 3$ matrix
multiplying ${\bf E}$, and we solve for the complex values
of $\omega$ that make the determinant of this matrix zero.
We use the numerical Newton-Raphson technique to isolate
individual solutions from a grid of starting guesses in
$\omega_r$, $\omega_i$ space.
Besides the assumption of Maxwellian distributions (see
below) and the use of numerical expansions for Bessel functions
and the plasma dispersion function (Fried \& Conte 1961;
Poppe \& Wijers 1990), the kinetic dispersion tensor
is calculated exactly.
The nine elements of $\widetilde{\mbox{\boldmath $\epsilon$}}$
at each solution allow us to compute the electric field
polarization vectors and the relative amounts of wave energy
in electric, magnetic, kinetic, and thermal perturbations for
each wave mode (see eq.~[\ref{eq:dUmax}] and Appendix 1 of
Krauss-Varban et al.\  1994).

We model the local plasma conditions at a heliocentric
distance of 2 $R_{\odot}$ in the accelerating high-speed
solar wind.
The magnetic field is given by the polar coronal-hole
flux tube model of Banaszkiewicz et al.\  (1998), with
$\Omega_{p} = 7660$ rad s$^{-1}$.
The coronal-hole electron density $n_e$ at 2 $R_{\odot}$
has been constrained to be between $2 \times 10^5$ and
$7 \times 10^5$ cm$^{-3}$, depending on the relative
concentration of polar plumes (Cranmer et al.\  1999b).
We choose a value within this range that sets the
Alfv\'{e}n speed $V_A$ to be exactly 2000 km s$^{-1}$.
For the purpose of computing the wave dispersion, we
assume isotropic Maxwellian distributions for the
protons and electrons with temperatures constrained
empirically as follows: $T_{e} = 8 \times 10^5$ K,
$T_{p} = 2 \times 10^6$ K (Kohl et al.\  1997,
1998; Esser et al.\  1999; Doschek et al.\  2001).
The plasma beta, here denoted as $\widetilde{\beta}$
to distinguish it from the cascade advection strength
$\beta$, is defined as the ratio of the total plasma
pressure to the magnetic pressure,
\begin{equation}
  \widetilde{\beta} \, \equiv \,
  \frac{8\pi}{B_{0}^2} \sum_{s} n_{s} k_{\rm B} T_{s}
  \, \approx \, 0.009
  \,\,\, ,
\end{equation}
where the numerical value pertains to the coronal
conditions described above.
At all heights in the corona, $\widetilde{\beta} \ll 1$,
but this quantity always remains larger than the
electron-to-proton mass ratio $m_{e} / m_{p}$.

In this paper we concentrate only on the solution branch
corresponding to shear Alfv\'{e}n waves.
This mode is expected to be dominant in the limit of
nearly incompressible MHD turbulence
(Goldstein et al.\  1995; Tu \& Marsch 1995),
and here we follow the evolution of Alfv\'{e}nic
fluctuations both to the high-frequency ion cyclotron
limit and to the high-$k_{\perp}$ KAW limit.
The Alfv\'{e}n mode is identified clearly at the lowest
wavenumbers in our computational grid
($k_{\para} V_{A} / \Omega_{p} = 10^{-7}$ and
$k_{\perp} R_{p} = 10^{-6}$), and it is tracked incrementally
as either $k_{\para}$ or $k_{\perp}$ is increased slowly.
We ignore the fast and slow magnetosonic wave
modes---and thus possibly underestimate the full
heating rates in the extended corona---but these types
of waves have long been thought to be damped substantially
before they can reach the corona.
We also ignore high-frequency plasma waves such as
Langmuir and hybrid modes (which are expected to be much
smaller in amplitude than the lower-frequency MHD waves)
as well as other strongly damped solutions with no
fluid counterparts (such as the Bernstein modes).

Figures 2 and 3 show example solutions of the dispersion
relation for various ranges of wavenumber.
Figure 2a plots the real and imaginary parts of the
frequency for a constant value of $k_{\para}$ (in the
low-frequency limit, $\omega_{r} \ll \Omega_p$) as
a function of $k_{\perp}$.
This figure thus illustrates the evolution of the KAW
in the limit of large obliqueness angles.
The ratio $| \omega_{i} / \omega_{r} |$ is negligibly small
for small values of $k_{\perp}$, but it grows to a value of
order 0.2 in the KAW limit ($k_{\perp} R_{p} \gtrsim 1$).
The damping for all plotted values of $k_{\perp}$ is
dominated by the linear Landau resonance.
Figure 2b plots the relative contributions to the total
linearized energy density of the waves as defined in
eq.~(\ref{eq:dUmax}).
In the low-$k_{\perp}$ MHD limit, there is equipartition
between the magnetic field term and the proton kinetic term
(i.e., transverse velocity oscillations).
When there is significant KAW dispersion, though, the
electron kinetic term becomes important as fewer protons
remain coupled to the small-scale fluctuations.
``Thermal'' density fluctuations also become non-negligible
(Hollweg 1999a).
The protons and electrons exhibit identical relative
density oscillations $\delta n / n$, but the proton thermal
energy density is larger because $T_{p} = 2.5 T_{e}$.
The quantity $\phi_e$ (eq.~[\ref{eq:phidef}]), which is
defined as the ratio of kinetic energy in perpendicular
electron fluctuations to the total energy density, is also
plotted in Figure 2a.

In the low-frequency, highly oblique KAW limit, the Alfv\'{e}n
branch ceases to exist for perpendicular wavenumbers greater
than $k_{\perp} R_{p} \approx 14$.
This occurs when the parallel phase speed of the waves
$\omega_{r} / k_{\para}$ exceeds approximately two times
the electron most-probable speed, and the number of electrons
available to participate in coherent wave motions (in the tail
of the distribution) becomes increasingly small.
We believe this also explains the departure from the simple
analytic KAW dispersion relation,
\begin{equation}
  \omega_{r} \, \approx \, k_{\para} V_{A}
  \sqrt{1 + k_{\perp}^{2} R_{p}^{2}}  \,\,\, ,
\end{equation}
where a temperature-dependent, order-unity factor
multiplying the $k_{\perp}^{2}$ term has been omitted.
The above expression assumes full coupling between the
waves and the entire velocity distribution.
The cessation of solutions occurs at higher perpendicular
wavenumbers for higher $\widetilde{\beta}$ plasmas.

For higher wavenumbers ($k_{\perp} R_{p} \gtrsim 14$) in
the low-$\widetilde{\beta}$ plasma considered here, the
kinetic extension of the fast-mode magnetosonic wave
is the only surviving solution with similar dispersion
properties as the KAW.
This mode is similarly longitudinal (i.e., with its
electric field polarization vector nearly parallel to its
wavevector), and it exhibits a similar distribution of energy
density fractions as in the rightmost part of Figure 2b.
Although we expect the turbulent cascade to deposit some
power into this new mode once the KAW solutions disappear,
we neglect this mode henceforth in order to determine
how much of the turbulent energy is absorbed by the purely
Alfv\'{e}nic fluctuations.

Figure 3 displays the dispersion properties of nearly
parallel-propagating Alfv\'{e}n waves versus $k_{\para}$
(with $k_{\perp} R_p$ held constant at $10^{-3}$) as the
wave frequency approaches the proton cyclotron frequency.
The numerically computed frequency comes close to that of
the cold-plasma approximation,
\begin{equation}
  \omega_{r} \, \approx \, k_{\para} V_{A}
  \sqrt{1 - \frac{\omega_r}{\Omega_p}}
\end{equation}
(e.g., Dusenbery \& Hollweg 1981).
The damping rate $\omega_i$ contains two contributions: one
due to Landau damping (at low $k_{\para}$) and one due to
the proton cyclotron resonance (at high $k_{\para}$).
The waves in this regime are nearly incompressible, with
the magnetic energy density varying from equipartition with
proton bulk motions in the MHD limit, to near dominance
of the total fluctuation energy in the limit
$\omega_{r} \rightarrow \Omega_p$ (see Figure 3b).

We find that the Alfv\'{e}nic proton cyclotron branch ceases
to exist for parallel wavenumbers greater than
$k_{\para} V_{A} / \Omega_{p} \approx 3.4$.
This ``warm plasma'' effect was predicted to occur by
Stix (1992) at about
\begin{equation}
  \frac{k_{\para {\rm max}} V_A}{\Omega_p} \, = \,
  \pi^{1/6} \, \left( \frac{V_A}{w_{\para p}} \right)^{1/3}
  \,\,\, ,
\end{equation}
where $w_{\para p}$ is the parallel proton most-probable speed,
and this quantity is $\sim$2.7 in the plasma conditions
modeled here.
The 20\% difference between the predicted and actual cutoff
values probably occurs because of the different assumed proton
and electron temperatures (electron contributions are ignored
in the above expression).

We computed the real and imaginary parts of the Alfv\'{e}n-mode
frequency $\omega$ in a grid that extended from $10^{-7}$ to
3.4 (in $k_{\para} V_{A} / \Omega_p$) and from $10^{-6}$ to
14.2 (in $k_{\perp} R_p$).
Specifically, $\omega_{i}$ is the growth or damping rate
(growth if positive; damping if negative) for the electric
field perturbation amplitude.
In order to determine the positive-definite damping rate
for the dominant two-dimensional cascade of energy
(proportional to the square of the amplitude), we set the
quantity $D$ in eq.~(\ref{eq:dWdt}) equal to $-2 \omega_i$.
We selected a locus of points in the wavenumber grid that
traces out the critical balance condition of $y=1$, with $y$
defined preliminarily by the {\em undamped} spectrum
$W_{\perp}$.
For most values of the wave power that we considered, the
$y=1$ curve occurs where the wavevectors are extremely
oblique ($k_{\perp} \gg k_{\para}$), and thus they are in
the KAW regime plotted in Figure 2.
For these low-frequency fluctuations, the ratio
$\omega_{i} / \omega_{r}$ depends only on $k_{\perp}$, and
we can write the frequency and the critically balanced damping
rate along the $y=1$ curve as
\begin{displaymath}
  \omega_{r} \, = \, k_{\perp} v_{\perp} \, = \,
  k_{\perp} \phi_{e}^{1/2} \, W_{\perp}^{1/2}  \,\,\, ,
\end{displaymath}
\begin{equation}
  D ( k_{\perp} ) \, = \, -2 \omega_{i} \, = \,
  - \frac{2\omega_i}{\omega_r}
  \, k_{\perp} \phi_{e}^{1/2} \, W_{\perp}^{1/2}
  \,\,\, .
\end{equation}
We found a parameterized fit for the $k_{\perp}$ dependence
of the ratio $\omega_{i} / \omega_{r}$ and combined it with
eq.~(\ref{eq:fedef}) to derive the following expression for
the damping rate along the $y=1$ curve:
\begin{equation}
  D ( k_{\perp} ) \, \approx \,
  (5.95 \times 10^{-5} \, \mbox{cm}^{-1})
  \frac{k_{\perp}^{3} R_{p}^{3} \, W_{\perp}^{1/2}}
  {(1 + 0.7 k_{\perp}^{2} R_{p}^{2})^{1/2}}
  \,\, .
  \label{eq:Dfit}
\end{equation}
The numerical factor on the right-hand side applies
specifically to the adopted coronal conditions at
2 $R_{\odot}$.
Also, this parameterization is valid only for the range of
wavenumber where there are Alfv\'{e}nic solutions (i.e.,
$k_{\perp} R_{p} \lesssim 14$).
In the MHD limit, where $k_{\perp} R_{p} \ll 1$,
$D (k_{\perp})$ is proportional to $k_{\perp}^{8/3}$, which
is close to---but slightly steeper than---the commonly assumed
viscous damping relation $D \propto k_{\perp}^2$.

\subsection{Quasi-linear Proton and Electron Heating}

The above calculation of the damping rate $\omega_i$ does
not reveal the relative amounts of energy absorbed by
protons or electrons, nor does it describe how the particle
velocity distributions are driven away from isotropy.
In this section we compute individual proton and electron
heating rates (in both the parallel and perpendicular
directions) and bulk acceleration rates, based on
assumptions utilized by Quataert (1998),
Quataert \& Gruzinov (1999), and Marsch \& Tu (2001).
The individual heating and acceleration rates are
computable in closed form in the quasi-linear limit
(i.e., if $| \omega_{i} | \ll \omega_r$).
We use this assumption to compute only the relative
fractions of energy in the various rates, and normalize
the total energy dissipation by the numerically computed
(not quasi-linear) values of $\omega_i$ from the previous
section.

The approach used in this paper differs from that of
Leamon et al.\  (1999), who isolated the electron and proton
contributions to the turbulent heating rate by comparing their
baseline calculation with a trial calculation with an extremely
low electron $\widetilde{\beta}$.
The approach of Leamon et al.\  retained the full solution of
the linear dispersion relation (even for
$| \omega_{i} | \approx \omega_r$), but at the expense of
having to assume an unrealistically small electron temperature.
Leamon et al.\  (1999) also evaluated the KAW dispersion
relation in plasma conditions appropriate for the {\em in situ}
solar wind ($r \gtrsim 20 \, R_{\odot}$), whereas we are
concerned with the corona.

To retain the greatest generality, we assume a bi-Maxwellian
velocity distribution for each species $s$,
\begin{displaymath}
  f_{0,s} (v_{\para}, v_{\perp}) \, = \,
  \frac{n_s}{\pi^{3/2} w_{\para s} w_{\perp s}^2}
\end{displaymath}
\begin{equation}
  \times \, \exp
  \left[ - \left( \frac{v_{\para} - u_{\para s}}{w_{\para s}}
  \right)^{2} - \frac{v_{\perp}^2}{w_{\perp s}^2} \right]
  \,\,\, ,
\end{equation}
where $u_{\para s}$ is the bulk speed along the magnetic field
and $w_{\para s}$ and $w_{\perp s}$ are the anisotropic
most-probable speeds.
These quantities relate to the temperatures via
\begin{equation}
  w_{\para s}^{2} \equiv 2 k_{\rm B} T_{\para s} / m_{s}
  \,\, , \,\,\,\,\,\,
  w_{\perp s}^{2} \equiv 2 k_{\rm B} T_{\perp s} / m_{s}
  \,\,\, .
\end{equation}
Let us define the damping rate that is attributable to
particle species $s$ as a positive quantity:
\begin{equation}
  \omega_{i,s} \, \equiv \, - \omega_{i} \, \left\{
  \frac{\omega_{ps}^{2} \,
  ( \omega_{r} - u_{\para s} k_{\para} ) \, K_{s}^{(0)}}
  {\sum_{t} \omega_{pt}^{2} \,
  ( \omega_{r} - u_{\para t} k_{\para} ) \, K_{t}^{(0)}}
  \right\}
  \label{eq:MTUomega}
\end{equation}
where $\omega_{ps} = (4\pi q_{s}^{2} n_{s}/m_{s})^{1/2}$
is the plasma frequency of species $s$ and $q_s$ is the
charge on particle $s$.
The denominator of eq.~(\ref{eq:MTUomega}) ensures that
the sum of $\omega_{i,s}$ over all species must be equal to
the magnitude of $\omega_i$.
We define the generalized resonance function $K_{s}^{(m)}$ as
\begin{equation}
  K_{s}^{(m)} \, \equiv \, \sum_{\ell = -\infty}^{+\infty} \,
  \ell^{m} \, {\cal R}_{s} ({\bf k}, \ell)
\end{equation}
where ${\cal R}_{s} ({\bf k}, \ell)$ is the linear
Landau/cyclotron resonance function given by eqs.~(42)--(45)
of Marsch \& Tu (2001) and $m$ is an integer.
The quantity ${\cal R}_s$ depends on complicated functions
of $\omega$, ${\bf k}$, and the electric field polarization
vector, and is too lengthy to reproduce here.
Linear wave-particle resonances are included in this
quantity as factors of the form
\begin{equation}
  \exp \left[ - \left(
  \frac{\omega - u_{\para s} k_{\para} - \ell \Omega_s}
  {w_{\para s} k_{\para}} \right)^{\! 2 \,} \right]
  \label{eq:resexp}
\end{equation}
where $\ell = 0$ corresponds to Landau damping and transit
time magnetic pumping (the latter believed to be important only
when $\widetilde{\beta} \gtrsim 1$), and $\ell \neq 0$
corresponds to cyclotron or gyroresonant damping.
In the models presented below, we compute ${\cal R}_s$ fully
for the Alfv\'{e}n modes discussed in the previous section,
and we sum $K_{s}^{(m)}$ over $\ell$ from $-$20 to $+$20.
Test runs with larger ranges of $\ell$ have ensured that
no quantitative results depended on this choice of lower
and upper summation bounds.

Marsch \& Tu (2001), following Melrose (1986) and others,
wrote the particle heating rates as an integration over
either a magnetic or an electric power spectrum.
We have rewritten these rates in terms of $W ({\bf k})$
and $\omega_{i,s}$ by noting that the total heating rate
for species $s$ ($Q_s$) should be balanced by the
damping of the total fluctuation spectrum, i.e.,
\begin{displaymath}
  Q_{s} \, = \, Q_{\para s} + Q_{\perp s} \, = \,
  n_{s} k_{\rm B} \frac{\partial}{\partial t}
  \left ( \frac{T_{\para s}}{2} + T_{\perp s} \right)
\end{displaymath}
\begin{equation}
  \, = \, \rho \int d^{3} {\bf k} \,\,
  W ({\bf k}) \,\, 2 \omega_{i,s}
  \,\,\, .
\end{equation}
The individual resonant acceleration and heating rates
are thus given by an analogue of eqs.~(27) and (33) of
Marsch \& Tu (2001):
\begin{displaymath}
  m_{s} n_{s} \frac{\partial}{\partial t} \left\{ \!
  \begin{array}{c}
    u_{\para s} \\
    w_{\para s}^{2} \\
    w_{\perp s}^{2}
  \end{array}
  \! \right\} \, \equiv \, \left\{ \!
  \begin{array}{c}
    A_{\para s} \\
    4 Q_{\para s} \\
    2 Q_{\perp s}
  \end{array}
  \! \right\} \, = \,
\end{displaymath}
\begin{displaymath}
  \rho \int d^{3} {\bf k} \,\,
  W ({\bf k}) \,
  \left( \frac{2 \omega_{i,s}}{\omega_{r} -
  u_{\para s} k_{\para}} \right) 
\end{displaymath}
\begin{equation}
  \times \, \left\{ \!
  \begin{array}{c}
    k_{\para} \\
    4 ( \omega_{r} - u_{\para s} k_{\para} ) -
    4 \Omega_{s} K_{s}^{(1)} / K_{s}^{(0)} \\
    2 \Omega_{s} K_{s}^{(1)} / K_{s}^{(0)}
  \end{array}
  \! \right\}  \,\, .
  \label{eq:MTUheat}
\end{equation}
For consistency with the assumptions in {\S}~3.1, we have
solved the above equations for isotropic Maxwellians
($w_{\para s} = w_{\perp s}$) and in the frame of
the solar wind's bulk outflow (i.e., $u_{\para s} = 0$).
The resonant acceleration term presented above
is given only for completeness.
It has been shown that $A_{\para s}$ is probably not an
important contributor to solar wind acceleration, since the
magnetic mirror force is typically stronger by a wide margin
(e.g., Hollweg 1999b, Cranmer 2001).

To evaluate eq.~(\ref{eq:MTUheat}), the wave spectrum
$W (k_{\para}, k_{\perp})$ must be specified on the same
grid of wavenumber points as in the dispersion calculation.
The two-dimensional wave spectrum $W_{\perp} (k_{\perp})$
was first computed with no damping in order to specify the
location of the $y=1$ curve, then it was recomputed with the
damping rate $D (k_{\perp}$) given in eq.~(\ref{eq:Dfit}).
The change in $W_{\perp}$ did not significantly affect the
position of the $y=1$ curve in wavenumber space, so further
iteration was not needed.
We used the full numerical solutions for $\omega_r$ and
$\phi_e$ in the definition of $y$ so that many of the
simplifying assumptions in {\S}~2.3 did not need to be applied.
However, the analytic expression for $g(y)$ in
eqs.~(\ref{eq:kapfun}) and (\ref{eq:enndef}) was used, with
$\zeta$ given by a simple undamped approximation that applies
in the MHD and KAW limits,
\begin{equation}
  \zeta \, \approx \, \frac{1}{3} \left(
  \frac{1 + 2 k_{\perp}^{2} R_{p}^2}
  {1 + k_{\perp}^{2} R_{p}^2} \right)
  \,\,\, .
\end{equation}
The parallel diffusion strength $\alpha$ was always assumed
to be equal to the perpendicular diffusion strength $\gamma$.
The analytic expression for $g(y)$ does not take account of
cyclotron damping at high values of $y$, which implies that the
computed proton heating rates should be considered upper limits.

Figure 4 displays various quantities in wavenumber space for
the baseline model that assumes $\alpha = \beta = \gamma = 1$
and $(\delta U / \rho)^{1/2} = 50$ km s$^{-1}$.
Contours of the proton and electron damping rates
(dimensionless ratios $| \omega_{i,s} / \omega_{r} |$,
for $s = p,e$) were computed in the same manner as the curves
in Figures 2 and 3, separated into proton and electron
contributions using eq.~(\ref{eq:MTUomega}), and plotted in
Figure 4a.
In Figure 4a we also give indications of the angle $\theta$
between the background magnetic field and the wavevector.
Note that significant proton cyclotron damping occurs (for
Maxwellian proton distributions) only when $k_{\para}$
exceeds the inverse inertial length $\Omega_{p} / V_A$,
with a rapid, Gaussian decline for smaller values of
$k_{\para}$.
The onset of electron Landau damping, on the other hand,
occurs much more gradually in the high-$k_{\perp}$ KAW limit.

Figure 4b shows contours of $W ({\bf k})$, computed as
described above, and the critical balance ($y=1$) curve.
Although the decay of fluctuation energy as a function of
$k_{\para}$ is steep, there is some residual energy that
seems to make it to the proton cyclotron frequency.
We should note, however, that for the case plotted in
Figure 4b the solutions for
$k_{\para} V_{A} / \Omega_{p} \gtrsim 10^{-2}$
are only approximate.
The explanation for the power-law behavior of $g(y)$
at large values of $y$ depends on the existence of
perpendicular wavenumber diffusion from high $k_{\perp}$
to low $k_{\perp}$.
As can be seen in Figure 4b, there is sufficient
Alfv\'{e}nic power available to be perpendicularly diffused 
only for $k_{\para} V_{A} / \Omega_{p} \lesssim 10^{-2}$
(i.e., to the right of the $y=1$ curve), and the extension
of the power law to larger values of $k_{\para}$ is an
assumption that needs to be investigated further.
We speculate that the inclusion of other (non-Alfv\'{e}n)
modes at higher values of $k_{\perp}$ could make more
power available to be diffused over a larger range of
$k_{\para}$.

The integration of eq.~(\ref{eq:MTUheat}) provides the
individual heating and acceleration rates.
Below, we express these rates (i.e., $Q_{\para s}$ and
$Q_{\perp s}$) dimensionlessly by dividing them by a
fiducial empirical value $Q_{\rm emp} \equiv 10^{-8}$
erg cm$^{-3}$ s$^{-1}$, which is representative of required
heating rates at 2 $R_{\odot}$ in various published models
(e.g., Leer et al.\  1982; Esser \& Habbal 1995;
Li et al.\  1999; Dmitruk et al.\  2002).
For the baseline model illustrated in Figure 4, the dominant
dissipation channel is Landau damping, which is converted
mainly into parallel electron heating.
In this model, $Q_{\para e} / Q_{\rm emp} = 0.26$, and
$Q_{\perp e}$ is identically zero.
There is a small amount of high-$k_{\para}$ cyclotron
energization, which heats protons in the perpendicular
direction ($Q_{\perp p} / Q_{\rm emp} = 9.1 \times 10^{-5}$)
and cools them slightly in the parallel direction
($Q_{\para p} / Q_{\rm emp} = -1.9 \times 10^{-5}$).
These rates, though, are far too small to provide the
required energy to protons in the extended corona.

We performed a number of rate calculations with different
values for $\beta$, $\gamma$, and $\delta U$.
The total heating rates scale closely with the injected
cascade rate $\varepsilon_{0} \sim (\delta U / \rho)^{3/2}$.
Depending on the ratio $\beta / \gamma$, though, as much as
30\% of the input cascade energy can be unaccounted
for in the total Alfv\'{e}n-wave damping.
This is not surprising because we cut off the wave spectrum
abruptly when the Alfv\'{e}n branch solutions disappear and
did not follow the flow of ``mode-coupled'' energy to higher
wavenumbers.
Following the complete turbulent cascade of energy on more
than one dispersion branch is an important subject for
future work.

The following phenomenological scaling relations were produced
as empirical fits to the numerical results found by varying
$\beta$, $\gamma$, and $\delta U$:
\begin{equation}
  \frac{Q_{\para e}}{Q_{\rm emp}} \, \approx \, 0.22 \,
  \left[ \frac{( \delta U / \rho )^{1/2}}{50 \, \mbox{km~s}^{-1}}
  \right]^{3} \left( \beta + \frac{2\gamma}{3} \right)^{1/3}
\end{equation}
\begin{equation}
  \frac{Q_{\perp p}}{Q_{\rm emp}} \, \approx \, 9.1 \times 10^{-5}
  \, \left[ \frac{( \delta U / \rho )^{1/2}}{50 \, \mbox{km~s}^{-1}}
  \right]^{2n+1} \, y_{\rm res}^{3.5 - 2n}
  \,\,\, .
  \label{eq:empQperp}
\end{equation}

\noindent
The electron heating rate is relatively insensitive
to the adopted values of $\beta$ and $\gamma$, and its absolute
value agrees with the empirically constrained total heating rate
(i.e., $Q_{\para e} = Q_{\rm emp}$) if
$(\delta U / \rho)^{1/2} \approx 120$ km s$^{-1}$.

Note the sensitive dependence of the proton heating rate on
the exponent $n$ in the power-law tail of $g(y)$.
In the MHD inertial range, $n$ is given
simply by $1 + (3 \beta / 4 \gamma)$.
The baseline case of $\beta = \gamma$ sets the normalization
in the exponent above (i.e., for this case $2n = 3.5$).
The dimensionless factor $y_{\rm res}$, which we constrained
by fitting to be about 770, gives a rough indication of how
far out in the tail the cyclotron interaction occurs.
For proton heating, the dominant contribution to the integral
in eq.~(\ref{eq:MTUheat}) occurs when
$k_{\para} \approx k_{\perp}$ (i.e., when
$k_{\perp} R_{p} \approx 0.2$ at the proton cyclotron
resonance) and at this wavenumber, $y \approx 10^3$.

Figure 5 plots the perpendicular proton heating rates from
eq.~(\ref{eq:empQperp}) versus the ratio $\beta / \gamma$.
Also plotted is a range of empirical proton heating rates
from Li et al.\  (1999) that produce agreement with
UVCS/{\em{SOHO}} \ion{H}{1} Ly$\alpha$ observations.
It is clear that if $\beta = \gamma$, one would need
unrealistically high values of the total wave power to agree
with the empirical rates.
However, the required levels of wave power are greatly
reduced if $\gamma > \beta$.
For example, Esser et al.\  (1999) derived an empirical
range of 110 to 130 km s$^{-1}$ for the total Alfv\'{e}n wave
velocity amplitude at 2 $R_{\odot}$, using wave action
conservation and spectroscopic limits on the base amplitude.
To reach the empirical range of proton heating rates in
Figure 5 simultaneously with this constraint on
$(\delta U / \rho)^{1/2}$, one would need $\gamma$ to be
greater than, or of the same order as, $\sim 4 \beta$.
This does not seem to be an unreasonable condition on the
turbulent cascade, but only simulations or laboratory
experiments can provide firm limits on how the effective
advection and diffusion strengths compare with one another.

The use of $\delta U$ as a total wave energy quantity glosses
over the fact that, at the outer scale, there is probably
considerably more power in outward-propagating waves
($k_{\para} > 0$) than in inward-propagating waves
($k_{\para} < 0$).
Let us compare the heating rates computed above with the
turbulence model of Matthaeus et al.\  (1999) and
Dmitruk et al.\  (2001, 2002), who took the asymmetry in
the outward and inward populations into account.
Their net heating rate can be expressed as
\begin{equation}
  Q \, \approx \, \rho \,
  \frac{Z_{-}^{2} Z_{+} + Z_{+}^{2} Z_{-}}{L_{\rm out}}
  \label{eq:dmit}
\end{equation}
where $Z_{-}$ and $Z_{+}$ are wavenumber-integrated Els\"{a}sser
amplitudes (with velocity units) of outward and inward
propagating fluctuations, respectively, and $L_{\rm out}$ is
a representative energy-containing length scale.
Dmitruk et al.\  (2001, 2002) also approximated the
reflected inward power $Z_{+}$ as
$\sim L_{\rm out} | \partial V_{A} / \partial r |$.
For the parameters at 2 $R_{\odot}$ discussed in {\S}~3.1,
the outer scale $L_{\rm out}$ assumed in our cascade equation,
and $Q = Q_{\rm emp}$, we solve for $Z_{-}$ and $Z_{+}$
to be approximately 30 and 7 km s$^{-1}$, respectively.
These quantities are amplitudes most comparable to the
peak value of $W_{\perp}^{1/2}$ at the outer scale, so in
order to compare them with $(\delta U / \rho)^{1/2}$ we
must multiply by the factor of 5 that arises because of the
broadening of $W_{\perp}$ as a function of $k_{\perp}$
(see {\S}~2.2).
This implies a total outward wave power equivalent to
$(\delta U / \rho)^{1/2} \approx 150$ km s$^{-1}$, which is
close to the empirical range reported by Esser et al.\  (1999).
Our requirement above that $(\delta U / \rho)^{1/2}$ must
be of order 120 km s$^{-1}$ is also within this range.
Thus, although we did not consider outward/inward asymmetries
explicitly, the derived requirements on the total wave power
seem robust and useful nonetheless.

It is still uncertain whether the dissipation of Alfv\'{e}nic
turbulence in the extended corona will preferentially heat
electrons or protons.
If van Ballegooijen's (1986) statistical limit of
$\beta = \gamma$ holds for actual strong MHD or KAW turbulence,
it seems clear that the electrons will be primarily heated
along the magnetic field and there would not be sufficient
wave power at the ion cyclotron frequencies to heat protons
and heavy ions.
This represents a problem when confronted with spectroscopic
observations of extreme ion heating and anisotropy (with
$T_{\perp} \gg T_{\para}$) and at least mild preferential
heating of protons over electrons (e.g.,
Kohl et al.\  1996, 1997, 1998; Esser et al.\  1999;
Cranmer et al.\  1999b; Doschek et al.\  2001).
We discuss alternate ways of energizing protons and heavy
ions in the extended corona below in {\S\S}~4--5.

\section{Nonlinear Evolution of Electron Beams}

Here we explore the possible consequences of the strong (but
largely unobserved) parallel electron heating that would be
the result of Landau damping of kinetic Alfv\'{e}n waves.
The ideas discussed in this section are considerably more
speculative than those discussed above, but we are guided by
detailed {\em in situ} observations of roughly similar
conditions in the Earth's ionosphere and magnetosphere.
We gratefully acknowledge Matthaeus et al.\  (2003a, 2003b)
for the initial suggestion that the following empirical
constraints in near-Earth plasma environments could be
applied to the solar corona and solar wind.

A possible sequence of events could progress as follows.
The Landau damping of Alfv\'{e}n waves heats electrons in
the parallel direction and has been shown, if allowed to
develop far enough, to lead to non-Maxwellian tails
(Tanaka, Sato, \& Hasegawa 1987, 1989; Leubner 2001).
These effectively beamed distributions may be unstable to
the direct generation of ion cyclotron waves
(Markovskii \& Hollweg 2002) which could heat protons.
However, we also note that electron velocity distributions
with monotonically decreasing tails may eventually develop
into non-monotonic bump-in-tail distributions if the
processes that accelerate electrons are intermittent or
otherwise spatially localized (e.g., Cairns 1987).
Even weak bump-in-tail electron beams have been shown to
easily generate parallel Langmuir waves via various plasma
instabilities (e.g., Thompson 1971; Melrose \& Goldman 1987;
Dum 1990; and references therein).
Evolved Langmuir wave trains exhibit a periodic electric
potential-well structure in which some of the beam electrons
can become trapped.
Adjacent potential wells can eventually merge with one
another and gradually form isolated ``holes'' of saturated
potential in electron phase space (Krasovsky, Matsumoto,
\& Omura 1999; Omura et al.\  2001; Mottez 2001).

Electron phase-space holes (EPHs) are equilibrium
electrostatic structures that have been predicted to
exist for some time (Bernstein, Greene, \& Kruskal 1957)
and have been observed in the last decade in space plasma
environments where electrons are accelerated along the
background magnetic field (e.g., Matsumoto et al.\  1994;
Ergun et al.\  1998, 1999; Bale et al.\  1998, 2002).
The traditional formation mechanism for EPHs has long
been considered to be the two-stream instability, but
the observed EPHs have potential amplitudes that are
too small to be formed from the strong counter-streaming
electron beams required to excite this instability.
The above bump-in-tail/Langmuir excitation scenario
is an alternate mechanism that has been shown to be able
to form EPHs in numerical simulations.
The stochastic nature of MHD turbulence (possibly
dissipating in small reconnection sites) has also been
shown to lead to similar kinds of intermittent
nonlinearities (Dupree 1982; Ambrosiano et al.\  1988;
Biglari \& Diamond 1989; Marsch, Tu, \& Rosenbauer 1996;
Tu, Marsch, \& Rosenbauer 1996; Drake et al.\  2003).

The reason we consider the effects of EPHs in the context
of this paper is that in many places where phase-space
holes are observed (e.g., Ergun et al.\  1998), protons
exhibit preferential heating in the direction perpendicular
to the magnetic field.
Ergun et al.\  (1999) suggested that EPHs act as
quasi-particles whose positively charged cores repel
ions in a manner similar to ion-ion Coulomb collisions.
Under certain conditions, a large number of these
stochastic collisions can lead to ion heating.
Because EPHs tend to flow along the magnetic field at
velocities of order the electron most-probable speed
($w_{\para e} \sim 5000$ km s$^{-1}$ in the extended
corona), the ions are roughly sitting still as the
EPHs flow by.
The net impulses the ions receive would thus be in the
perpendicular direction.
Below, we work out a quantitative estimate for the
perpendicular heating experienced by protons in the
extended corona due to EPH quasi-collisions.

\subsection{Phase Space Hole Properties}

Consider a homogeneous proton-electron plasma with the
following localized disturbance in the electrostatic
potential:
\begin{equation}
  \Phi (x,y,z) \, = \, \Phi_{0} \, \exp \left[
  - \left( \frac{z}{z_0} \right)^{2}
  - \left( \frac{r}{r_0} \right)^{2} \right]
  \,\,\, ,
  \label{eq:Phi}
\end{equation}
where the $z$ direction is that of the background magnetic
field and $r^{2} = x^{2} + y^{2}$.
The equipotential surfaces are oblate spheroids
centered around the origin, and the above distribution
is in agreement with several analyses of isolated EPH
properties (e.g., Turikov 1984; Chen \& Parks 2002).
The parallel length scale $z_0$ has been observed in several
space plasmas to be of the same order of magnitude as the
electron Debye length, and we define
\begin{equation}
  z_{0} \, \equiv \, \frac{\delta \, \lambda_D}{2}
\end{equation}
where $\lambda_D = w_{\para e} / \omega_{pe} \sqrt{2}$
is the electron Debye length and $\delta$ is a dimensionless
number that is observed to be typically 2 to 3.
(Ergun et al.\  1998).
The transverse length scale $r_0$ may scale with the
proton gyroradius in regions where
$\Omega_{e} / \omega_{pe} \gg 1$, but
Franz et al.\  (2000) estimated that
\begin{equation}
  \frac{r_0}{z_0} \, \approx \, \left(
  1 + \frac{\omega_{pe}^2}{\Omega_{e}^2} \right)^{1/2}
\end{equation}
in regions where $\Omega_{e} / \omega_{pe} \ll 1$ (which
is more applicable to the solar corona).
The above relation, which we apply in the model below, gives
an aspect ratio $r_{0} / z_{0}$ of about 3 to 10 in the
extended corona.

The localized electric field resulting from the above
potential can be determined analytically
(${\bf E} = - \nabla \Phi$), and the charge density
distribution is found by applying Coulomb's law.
The full analytic expression for the charge density
$\rho^{\ast} (r,z)$ agrees with the plots in Figure 3b
of Ergun et al.\  (1998), who modeled two extreme examples
of EPH geometry by effectively assuming $r_{0} = z_{0}$
(``sphere'') and $r_{0} \gg z_{0}$ (``plane'').
There is a net positive charge density in the central core
of the EPH which is surrounded by a net negative charge
density.
The positive charge density at the center of the EPH (i.e.,
at $r=0$, $z=0$) is expressible in terms of the difference
between local proton and electron number densities,
\begin{equation}
  \delta n \, \equiv \, n_{p} - n_{e} \, = \,
  \frac{\Phi_0}{2 \pi e} \left( \frac{2}{r_{0}^2}
  + \frac{1}{z_{0}^2} \right)
  \,\,\, ,
\end{equation}
which can be simplified further by neglecting the first
term in the parentheses (because $r_{0} \gg z_{0}$).
In this limit, we can express the ratio of $\delta n$ to
the total electron density $n_e$ as
\begin{equation}
  \frac{\delta n}{n_e} \, = \,
  \frac{16}{\delta^2} \,
  \frac{(e \Phi_{0} / m_e)}{w_{\para e}^2}
  \,\,\, .
  \label{eq:dnondef}
\end{equation}
This allows us to convert between the central potential
$\Phi_0$ and the central charge separation $\delta n$.

We assume the EPH moves along the $z$ direction with speed
$v_{\rm H}$.  This speed will probably be of the same order
as the parallel electron thermal speed, so let us follow
Turikov (1984) and define the Mach number
$M \equiv v_{\rm H} / w_{\para e}$.
Turikov (1984) found that one-dimensional EPHs become unstable
and break up when $M > 2$, and Ergun et al.\  (1998) report
values of $M$ between $\sim$0.2 and 2 in the ionosphere.
One cannot arbitrarily choose all three
of the quantities $\delta$, $M$, and $\delta n / n_e$.
They are interrelated by the constraint that the electrons
obey the Vlasov equation and can self-consistently affect
the electric potential via Poisson's equation.
Turikov (1984) computed a relation that allows us to solve
for $\delta n / n_e$ as a function of $\delta$ and $M$.
For a Gaussian potential in the $z$ direction
(eq.~[\ref{eq:Phi}]) and a weak charge separation
(i.e., $\delta n / n_{e} \ll 1$, which always seems to hold
for reasonable values of the parameters), 
\begin{equation}
  \frac{\delta n}{n_e} \, \approx \,
  \left( \frac{\delta}{\delta_0} \right)^{4} \, e^{-2 M^2}
  \label{eq:turikov}
\end{equation}
where $\delta_{0} = 4 \pi^{-1/4} (2 \ln 4 - 1)^{1/2}$ for the
Gaussian potential.
Coincidentally, this value of $\delta_0$ is extremely close
to a value of 4.0.
Chen \& Parks (2001) noted that the above relation is formally
only an upper limit on $\delta n / n_e$, but it is a useful
scaling relation that we will assume holds for EPHs in
the corona.
Note from eqs.~(\ref{eq:dnondef}) and (\ref{eq:turikov})
that the central potential $\Phi_0$ is proportional to
$\delta^6$, and thus is extremely sensitive to the
parallel size of the EPH.

One additional quantity that needs to be constrained is the
spatial number density of EPHs---or equivalently, their
filling factor---in the plasma.
Only a detailed analysis of EPH formation will yield
constraints on how frequently they appear in space after
significant merging has occurred, so here we use empirical
limits from the {\em in situ} measurements.
Defining the volume of an EPH as the equipotential surface
that is $1/e$ times its central potential, we relate their
spatial number density $n_{\rm H}$ to their fractional
filling factor $f_{\rm H}$ via
\begin{equation}
  n_{\rm H} \, = \, \frac{f_{\rm H}}{4\pi z_{0} r_{0}^2 / 3}
  \,\,\, .
\end{equation}
In a field of highly oblate EPH structures moving in the
$z$ direction, $f_{\rm H}$ can be estimated by analyzing the
mean time (or distance) between significant electric-field
``kicks'' felt by a slowly moving observer.
For EPHs distributed randomly in three-dimensional space,
$f_{\rm H}$ should be close to the ratio
$z_{0} / \Delta z$, where the denominator is the mean distance
between strong EPHs in the $z$ direction.
Ergun et al.\  (1998) found that there are often times when
$\Delta z \approx v_{\rm H} / \Omega_{p}$, which we use in
the rough calculations below.
The specification for the filling factor also must take
account of electron-electron Coulomb collisions, because if
these collisions occur on a time scale {\em faster} than
the time it takes to Landau-damp the MHD turbulence and build
up electron beams, then EPHs should not form at all.
Thus, let us modify the purely geometric filling factor by
an approximate collision term, with
\begin{equation}
  f_{\rm H} \, \approx \, \frac{z_{0} \Omega_p}{v_{\rm H}}
  \, \frac{1}{1 + ( \tau_{\rm diss} / \tau_{\rm coll} )^2}
  \label{eq:spacing}
\end{equation}
where $\tau_{\rm diss}$ is a turbulence dissipation time
scale that we approximate as $L_{\rm out} / Z_{-}$
(eq.~[\ref{eq:dmit}]) and is about 200~s for
$(\delta U / \rho)^{1/2} \approx 100$ km s$^{-1}$
(see {\S}~3.2).
The standard electron self-collision time is
\begin{equation}
  \tau_{\rm coll} \, \approx \, \frac{0.266 \, T_{e}^{3/2}}
  {n_{e} \ln \Lambda}
  \label{eq:taucoll}
\end{equation}
(Spitzer 1962), where we assume $\ln \Lambda = 21$.
When $\tau_{\rm coll} \ll \tau_{\rm diss}$ the EPH
filling factor approaches zero, and when
$\tau_{\rm coll} \gg \tau_{\rm diss}$ the filling factor
approaches the value constrained by Ergun et al.\  (1998).
For plasma conditions at 2 $R_{\odot}$, we can use canonical
values of $\delta = 3$ and $M=1$ to estimate $f_{\rm H}$
to be $\sim$10$^{-4}$, implying a spatial number density of
$3 \times 10^{-10}$ cm$^{-3}$.

\subsection{Quasi-Collisions and Proton Heating}

A single electron phase-space hole moving along the $z$ direction
encounters a positive ion (with mass $m_i$ and charge $q_i$)
at a perpendicular impact parameter, or distance of closest
approach, denoted by $r$.
We ignore Lorentz forces along the $z$ direction because they
are expected to roughly cancel out between the approaching
and receding halves of the EPH trajectory.
The perpendicular momentum given to the ion over the entire
trajectory is given by
\begin{equation}
  m_{i} \, \Delta v_{\perp i} \, = \, \int_{-\infty}^{+\infty}
  dt \, q_{i} \, E_{\perp} (t)
  \,\,\, ,
\end{equation}
where the time $t$ and the relative parallel displacement $z$
between the EPH and the ion are related simply by
$z = v_{\rm H} \, t$.
The perpendicular electric field felt by the ion is that of
the isolated EPH potential,
\begin{displaymath}
  E_{\perp} \, = \, ( E_{x}^{2} + E_{y}^{2} )^{1/2}
\end{displaymath}
\begin{equation}
  = \, \frac{2 r \Phi_0}{r_{0}^2} \, \exp \left[
  - \left( \frac{v_{\rm H} t}{z_0} \right)^{2}
  - \left( \frac{r}{r_0} \right)^{2} \right]
  \label{eq:Eperp}
  \,\,\, .
\end{equation}
We thus compute the perpendicular velocity impulse
analytically to be
\begin{equation}
  \Delta v_{\perp i} \, = \,
  \frac{2 \, \pi^{1/2} q_{i} \, r \, \Phi_{0} \, z_0}
  {m_{i} r_{0}^{2} \, v_{\rm H}} \, \exp \left[ -
  \frac{r^2}{r_{0}^2} \right]  \,\,\, .
\end{equation}
Following the standard development of Coulomb collision rates 
(e.g., Spitzer 1962; Jackson 1975), we extend the above
calculation from one EPH to a ``field'' of EPHs distributed
throughout space with number density $n_{\rm H}$.
The number of ion/EPH encounters per unit time with impact
parameters between $r$ and $r + dr$ is given by
$(2 \pi r \, dr \, v_{\rm H} n_{\rm H})$.
The effective {\em diffusion coefficient} for perpendicular
energization of the ion by the field of EPHs is then given by
\begin{displaymath}
  {\cal D}_{\perp i} \, \equiv \,
  \frac{d ( \Delta v_{\perp i}^{2} )}{dt} \, = \,
  \int_{0}^{\infty} dr \, 2 \pi r \, v_{\rm H} \, n_{\rm H}
  \, ( \Delta v_{\perp i} )^{2}
\end{displaymath}
\begin{equation}
  \, = \, \frac{\pi^{2} n_{\rm H} \, q_{i}^{2} \, \Phi_{0}^{2}
  \, z_{0}^2}{m_{i}^{2} \, v_{\rm H}}  \,\,\, .
  \label{eq:calD}
\end{equation}
Because of the exponential dependence of $E_{\perp}$ as a
function of $r$, the above integral was well-defined and finite
over all values of the impact parameter.
This is a relative simplification over the ion-ion Coulomb
collision problem, where the above integral formally diverges
and must be truncated using a maximum impact parameter.

In the limit of a field of very low-energy ions,
${\cal D}_{\perp i}$ is essentially the heating rate per
particle, per unit mass.
As the ions heat up, however, the amount of energy they
extract from the EPHs cannot be arbitrarily large.
Ideally, we should solve coupled energy conservation
equations for both ions and EPHs, where each equation
would contain energy exchange terms that would eventually
lead to equipartition.
It is not clear, though, how much of the EPH energy is
actually ``disposable'' (i.e., easily given away to ions
in collision-like interactions).

In the frame at rest with respect to a phase-space hole,
its electrostatic potential energy is given by the following
volume integration,
\begin{equation}
  U_{\rm E} \, = \, \int d^{3} {\bf x} \,
  \frac{| {\bf E} |^2}{8\pi}
  \,\, = \,\, \frac{\Phi_{0}^2}{8} \sqrt{\frac{\pi}{2}} \,
  \left( \frac{r_{0}^2}{2 z_0} + z_{0} \right)
  \,\,\, ,
\end{equation}
which is considerably larger than the kinetic energies of
individual protons or electrons in the solar corona.
In the frame of an ion, the EPH is moving by at a relatively
high speed (though we assume $v_{\rm H} \ll c$), and an
induced perpendicular magnetic field is felt by the ion
with a potential energy given by
\begin{equation}
  U_{\rm B} \, = \, \frac{v_{\rm H}^2}{c^2} \int
  d^{3} {\bf x} \, \frac{| E_{\perp} |^2}{8\pi}
  \,\, = \,\, \frac{\Phi_{0}^2}{8} \sqrt{\frac{\pi}{2}} \,\,
  \frac{z_{0} \, v_{\rm H}^2}{c^2}
  \label{eq:WH}
\end{equation}
which is several orders of magnitude smaller than the total
electrostatic energy of the EPH.
The induced magnetic energy is proportional to $v_{\rm H}^2$
and seems to be analogous to a kind of kinetic energy.
Our conjecture is that $U_{\rm B}$ represents an
order-of-magnitude estimate for the energy available to be
given away by EPHs in collisions with ions.
We express this energy quantity as an equivalent
temperature ($U_{\rm B} \equiv k_{\rm B} T_{\rm B}$) and
insert the adopted conditions at 2 $R_{\odot}$ (i.e.,
$\lambda_{D} = 14$ cm and $w_{\para e} = 5700$ km s$^{-1}$)
to obtain
\begin{equation}
  T_{\rm B} \, \approx \, ( 4.3 \times 10^{12} \mbox{K} )
  \left( \frac{\delta}{\delta_0} \right)^{13} M^{2}
  e^{-4 M^2}
\end{equation}
(see, e.g., eqs.~[\ref{eq:dnondef}] and [\ref{eq:turikov}]).
For increasing $M$, the total magnetic energy in the
EPH begins to decrease above about $M \approx 0.2$
because very fast EPHs contain fewer trapped electrons and
thus cannot support as high a potential as a slower EPH.

In the extended corona, the value of $T_{\rm B}$ is huge
in comparison to individual particle temperatures.
Further, this large value of $T_{\rm B}$ is {\em smaller} by
at least a factor of $v_{\rm H}^{2} / c^{2}$ than the
temperature quantity equivalent to the total electrostatic
energy of a phase-space hole.
EPHs seem to embody a substantial ``pool'' of energy that is
continually replenished by KAW turbulence and Landau damping,
and this energy does not seem to be diminished significantly
by collisions with ions.
Thus, we do not consider the approach to energy equipartition
any further and assume that eq.~(\ref{eq:calD}) gives the
effective net ion heating rate per unit mass.

For the specific case of protons, we solve a steady-state
solar wind internal energy conservation equation with the
EPH diffusion coefficient as the sole source of extended
heating.
Our approach ignores the impact of heat conduction and
energy sharing via particle-particle collisions, but these
effects are expected to be relatively weak in the extended
corona (e.g., Olsen \& Leer 1996; Li 1999).
We thus assume that the proton perpendicular temperature
$T_{\perp p}$ evolves with radius according to
\begin{equation}
  u_{\para p} \left( \frac{\partial T_{\perp p}}{\partial r}
  + \frac{T_{\perp p}}{A}
  \frac{\partial A}{\partial r} \right)
  \, = \, \frac{m_{p} {\cal D}_{\perp p}}{2k_{\rm B}}
  \,\,\, ,
  \label{eq:dTdr}
\end{equation}
where $u_{\para p} (r)$ is the time-steady outflow speed
profile and $A(r)$ is the cross-sectional area of a solar
wind flow tube.
The second term in parentheses above is responsible for
adiabatic proton cooling due to the geometric expansion of
the plasma.
Empirically determined values for the radial dependences
of $u_{\para p}$, $A$, $n_e$, $B_0$ (the background magnetic
field strength), and $T_e$ (the electron temperature; needed
to compute $\lambda_D$) are elaborated by Cranmer, Field,
\& Kohl (1999a).
We integrate eq.~(\ref{eq:dTdr}) from an inner boundary
at the base of the corona to an outer boundary in the
extended corona.
The remaining free parameters are the inner boundary value
of $T_{\perp p}$ and the EPH parameters $\delta$ and $M$,
which we assume to be constant in the extended corona.
We assume the EPH filling factor $f_{\rm H}$ is given by
eq.~(\ref{eq:spacing}).

Figure 6a shows the effective heating rate
$Q_{\perp p} \equiv \rho {\cal D}_{\perp p} / 2$,
computed using eqs.~(\ref{eq:dnondef}--\ref{eq:taucoll})
and eq.~(\ref{eq:calD}),
for various values of $\delta$ and a fixed value of $M=1$.
It also plots $Q$ for the Dmitruk et al.\  (2001, 2002)
turbulence dissipation rate (eq.~[\ref{eq:dmit}]), with the
radial dependence $\rho Z_{-}^{2} \propto V_{A} ( u_{\para p}
+ V_{A} )^{-2} A^{-1}$ given by wave action conservation
(see also Zank, Matthaeus, \& Smith 1996).
We interpret the order-of-magnitude similarity of the two
sets of rate quantities as a promising plausibility argument
for EPH proton heating.
Despite this similarity, though, the two sets of rates are
essentially ``apples and oranges,'' since the
Dmitruk et al.\  values come solely from the turbulent
cascade rate and the EPH values do not contain any
information about the cascade.
The EPH curves in Figure 6a are not energetically
consistent with the turbulence because we use the empirical
filling factor $f_{\rm H}$ discussed above.
A more self-consistent model of EPH heating would follow
the turbulent energy budget from the dissipation by Landau
damping to the growth of electron beams (which would embody
only a fraction of the total dissipated energy) to the
eventual intermittent formation of EPHs.

For protons that start out with $T_{\perp p} = 10^{6}$~K
at $r = 1.25 \, R_{\odot}$ (assuming some collisional contact
with electrons), Figure 6b plots the computed radial
dependence of $T_{\perp p}$ for the same range of $\delta$
values as in Figure 6a.
(The temperature curves are similar to those computed for
a single value of $\delta$ and a range of $M$ values.)
Substantial proton heating is evident, but the results are
very sensitive to the values of the EPH parameters.
It would probably be more realistic to assume distributions
of $\delta$ and $M$ (see, e.g., Figure 4a of
Ergun et al.\  1998), but this introduces free parameters
as well.
Note, though, that the mean value of $\delta$ reported by
Ergun et al.\  (1998) is $\sim$3.6, which is closest to
the value that would give $T_{\perp p} (r)$ most similar to
the UVCS \ion{H}{1} empirical model results
(Kohl et al.\  1998).

The proton heating models shown in Figure 6 are suggestive
but nowhere near conclusive.
A major source of uncertainty is the application of the
ionospheric EPH spacing relation
$\Delta z = v_{\rm H} / \Omega_p$ to the calculation
of the EPH filling factor in the corona.
This could be different by orders of magnitude, which would
significantly change the above results.
Also, our assumptions that $\delta$ and $M$ are constant with
radius and that eq.~(\ref{eq:turikov}) gives the exact
central EPH potential should be examined critically.
More work should be devoted toward investigating the
growth and development of EPHs in the corona, as well as
their interactions with protons, heavy ions, and the
``untrapped'' electrons.

\section{Discussion and Conclusions}

In this paper we suggested two alternative ``channels'' by
which protons and other positive ions in the extended solar
corona can be heated perpendicularly to the background
magnetic field (see Figure 7).
First, the parallel decay of anisotropic MHD turbulence to
ion cyclotron frequencies may occur if the cascade is
properly modeled by advection and diffusion strengths
$\beta$ and $\gamma$ with the property
$\gamma \gtrsim 4 \beta$.
This result could lend some support to the many other
proposed models that describe the damping of
parallel-propagating ion cyclotron waves in a coronal
context.
(Note, however, that many more issues, such as implementing
a more self-consistent cascade and damping formalism at
large values of $k_{\para}$, still needs to be resolved).
Second, we proposed that, if $\gamma < 4 \beta$, the
dominant parallel electron heating associated with KAW
Landau damping could lead to the nonlinear generation
of electron phase-space holes.
These EPHs have been shown to undergo collision-like
interactions with positive ions and heat them mainly
perpendicularly.
The EPH idea is extremely speculative because as yet we
have no quantitative constraint on their spatial filling
factor in the corona.
Only if they are generated in sufficient numbers (i.e.,
only if the conversion from KAWs to electron beams to
EPHs is efficient enough) can they provide substantial
proton heating.

There are several other related mechanisms by which protons
and heavy ions could be heated in the extended corona.
If there is substantial power in obliquely propagating
fast-mode magnetosonic waves, their collisionless damping
may contribute to proton heating (e.g., Li \& Habbal 2001;
Hollweg \& Markovskii 2002).
Oblique Alfv\'{e}n and fast-mode waves can steepen into
shocks under certain conditions, and numerical simulations
that employ the derivative nonlinear Schroedinger
equation (DNLS) have produced a rich variety of steepening
phenomena that produce power at high-frequency harmonics
of an input spectrum (e.g., Spangler 1997).
Certain types of collisionless shocks may also accelerate
ions in the direction perpendicular to the magnetic field
(Lee \& Wu 2000).
The basal generation and ``sweeping'' of high-frequency
waves (e.g., Axford \& McKenzie 1992; Tu \& Marsch 1997)
must also be developed further to determine to what degree
these waves can survive in the extended corona (see also
Cranmer 2001).
Because of the multiplicity of time and spatial scales
involved in studying many of these processes, the dominant
physics may become evident only when a large number of
competing mechanisms are included together in a solar wind
model and allowed to evolve self-consistently.

The proton heating that arises from ion cyclotron wave
dissipation depends sensitively on the shape of the proton
velocity distribution function.
We assumed Maxwellian distributions in the above analysis,
and the conclusions are not affected strongly if there is
moderate bi-Maxwellian anisotropy.
If the protons have suprathermal high-energy tails (e.g.,
Scudder 1992), though, the resonance function ${\cal R}_s$
would have a less steep frequency dependence compared to
eq.~(\ref{eq:resexp}).
In this case, waves with lower $k_{\para}$ would be included
in the thermally broadened ion cyclotron resonance and thus
lead to a a lower effective value of $y_{\rm res}$ in
eq.~(\ref{eq:empQperp}).
For strong enough suprathermal tails, it is possible that
even the baseline cascade model (with $\beta = \gamma$)
could contain enough parallel decay of wave energy to
heat protons substantially.
It should be noted, however, that Isenberg (2003) concluded
from a completely collisionless description of
``kinetic shell'' velocity distributions that the damping of
parallel-propagating ion cyclotron waves is not a likely
source of proton heating even if there is sufficient wave
power in the extended corona.
Indeed, damping rates for the non-Maxwellian distributions
predicted in quasi-linear models of collisionless resonant
heating can be substantially different than those computed
for Maxwellians (e.g., Isenberg 2001; Cranmer 2001; Tu \&
Marsch 2002).
Full kinetic models that follow the protons and electrons
from the collisional low corona to the collisionless extended
corona are needed to resolve these issues.

Some mention should be made of the effects of heavy ions
on the above analysis.
It has been clear for several decades that the 5\% to 10\%
abundance of He$^{2+}$ in the corona is responsible for a
strong resonance in the Alfv\'{e}n wave dispersion relation,
but it is unclear whether the other thousands of minor ion
species are numerous enough to affect the dispersion relation
(Isenberg 1984; Gomberoff, Gratton, \& Gnavi 1996).
Mode coupling between the different branches must be modeled
explicitly in order to determine how much wave power survives
at the proton cyclotron resonance.
Also, it should be made clear that the EPH quasi-collision
mechanism proposed above is {\em not} an efficient source of
preferential heating for heavy ions, even if it is able to
supply sufficient heat to the protons.
The effective ion heating rate is proportional to
$m_{i} {\cal D}_{\perp i}$, which scales with charge and
mass as $q_{i}^{2} / m_{i}$ (eq.~[\ref{eq:calD}]).
The O$^{5+}$ ions measured by UVCS/{\em{SOHO}} in the
extended corona have a perpendicular kinetic temperature
about two orders of magnitude larger than that of protons.
The EPH mechanism, though, would only provide an oxygen
temperature larger than the proton temperature by a
factor of $5^{2}/16 \sim 1.6$.
The damping of ion cyclotron waves still seems to be the
most likely source of preferential heating and acceleration
for heavy ions.

Improvements in remote-sensing measurements of the extended
solar corona are needed to make significant further progress
in identifying and characterizing the most important physical
processes.
Next-generation spectroscopic instruments are being designed
with the capability to sample the velocity distributions of
dozens of ions in the acceleration region of the high-speed
wind, as opposed to just 2--3 ions with UVCS.
In addition to being sensitive to many more emission lines,
these instruments could also detect subtle departures from
Gaussian line shapes that constrain the presence of specific
non-Maxwellian distributions (e.g., Cranmer 1998, 2001).
Seldom-used diagnostics such as the measurement of the
Thomson scattered \ion{H}{1} Ly$\alpha$ profile (which
probes the line-of-sight electron velocity distribution;
see Withbroe et al.\  1982) can put firm constraints on
both the ``core'' electron temperature and the existence
of beam-like or power-law wings.
Improvements in radio sounding observations are also making
it possible to extract a great deal of information about
MHD turbulence in the solar wind (e.g., Spangler 2002).

\acknowledgments
\noindent
The authors would like to thank William Matthaeus,
Jack Scudder, George Field, and S.\  Peter Gary
for valuable and inspiring discussions.
This work is supported by the National Aeronautics and Space
Administration under grants NAG5-11913 and NAG5-11420 to the
Smithsonian Astrophysical Observatory, by Agenzia Spaziale
Italiana, and by the Swiss contribution to the ESA PRODEX
program.

\end{multicols}

\appendix

\section{The Advection-Diffusion Cascade Model}

\small
This Appendix provides details concerning the solution of
eqs.~(\ref{eq:dWdt}) and (\ref{eq:dEdt}).
The two-dimensional advection-diffusion cascade model
(eq.~[\ref{eq:dWdt}]) can be simplified by defining the
following supplementary variables:
\begin{equation}
  \widetilde{W}(k_{\perp}) \, \equiv \, W_{\perp}^{3/2} \,
  k_{\perp}^{-3\beta / 2\gamma}
\end{equation}
\begin{equation}
  F (k_{\perp}) \, \equiv \, \frac{2\gamma}{3} \,
  \phi_{e}^{1/2} k_{\perp}^{2 + (3\beta / 2\gamma)}
  \,\,\, .
\end{equation}
We thus can define the perpendicular cascade rate
$\varepsilon$ as
\begin{equation}
  \varepsilon (k_{\perp}) \, \equiv \,
  - F (k_{\perp}) \frac{\partial \widetilde{W}}
  {\partial k_{\perp}} \,\,\, ,
\end{equation}
which in the MHD regime ($\phi_{e} = 1$) is proportional
to $k_{\perp} v_{\perp}^3$.
With the above definitions, eq.~(\ref{eq:dWdt})
simplifies to
\begin{equation}
  \frac{\partial W_{\perp}}{\partial t} \, = \,
  -k_{\perp} \frac{\partial \varepsilon}{\partial k_{\perp}}
  \, + \, S ( k_{\perp} ) \, - \, D ( k_{\perp} ) W_{\perp}
  \,\,\, .
  \label{eq:dQdt}
\end{equation}
In the inertial range, where both $S$ and $D$ are negligibly
small, it is easy to see that the time-steady solution exhibits
a constant value of $\varepsilon$, which we will denote
$\varepsilon_0$.
This cascade rate is supplied by the source function
$S (k_{\perp})$, which we estimate to have the following
Gaussian shape:
\begin{equation}
  S(x) \, = \, \frac{\varepsilon_0}{\pi^{1/2} \sigma_{\rm out}}
  \exp \left[ - \left( \frac{x - x_{\rm out}}{\sigma_{\rm out}}
  \right)^{2} \right]  \,\,\, ,
\end{equation}
where $x = \ln k_{\perp}$ and $x_{\rm out} = \ln k_{\rm out}$.
The dimensionless width of the Gaussian is given by
$\sigma_{\rm out} = 1$ in all models presented in this paper.
With this explicit form for $S$, the wavenumber dependence of
the time-steady cascade rate (at wavenumbers where $D \approx 0$)
can be found by integrating eq.~(\ref{eq:dQdt}), with
\begin{equation}
  \varepsilon (x) \, = \, \frac{\varepsilon_0}{2}
  \left[ 1 + \mbox{erf} \left(
  \frac{x - x_{\rm out}}{\sigma_{\rm out}} \right) \right]
  \,\,\, .
\end{equation}
This provides a smooth transition from zero cascade (when
$k_{\perp} \ll k_{\rm out}$) to cascade at the specified
rate $\varepsilon_0$ (when $k_{\perp} \gg k_{\rm out}$).
Thus, the solution for $\widetilde{W}$ in the inertial
range can be computed straightforwardly via
\begin{equation}
  \widetilde{W} ( k_{\perp} ) \, = \,
  \int_{k_{\perp}}^{+\infty}
  \frac{\varepsilon (k'_{\perp}) \, dk'_{\perp}}
  {F(k'_{\perp})} \,\,\, .
  \label{eq:Qquad}
\end{equation}
When $k_{\perp} \ll k_{\rm out}$, the cascade rate approaches
zero and $\widetilde{W}$ is constant, implying an energy
spectrum $W_{\perp} \propto k_{\perp}^{\beta / \gamma}$.
In the MHD inertial range, $\varepsilon$ is equal to a
constant value of $\varepsilon_0$ and eq.~(\ref{eq:Qquad})
can be integrated analytically to yield that $W_{\perp}^{3/2}$
is proportional to $k_{\perp}^{-1}$.
This results in the Kolmogorov-Obukhov spectrum
$W_{\perp} \propto k_{\perp}^{-2/3}$.
In the KAW inertial range, $\phi_{e} \propto k_{\perp}^2$
and eq.~(\ref{eq:Qquad}) integrates to obtain
$W_{\perp} \propto k_{\perp}^{-4/3}$.

For the full range of $k_{\perp}$ values below the dissipation
range, the integration of eq.~(\ref{eq:Qquad}) was performed
numerically to obtain the undamped curve in Figure 1.
For $\phi_e$ given by eq.~(\ref{eq:fedef}), an approximate
``bridging solution'' for the inertial range is
\begin{equation}
  W_{\perp} (k_{\perp}) \, \approx \, \left(
  \frac{3 \varepsilon_0}{2 \gamma \, k_{\perp}} \right)^{2/3}
  \left[ \left( 1 + \frac{3\beta}{2\gamma} \right)
  \sqrt{1 + k_{\perp}^{2} R_{p}^2} \, + \, k_{\perp} R_{p}
  \right]^{-2/3}
\end{equation}
which is valid in the MHD and KAW inertial-range limits
and differs by no more than 6\% from the exact numerical
solution when $k_{\perp} R_{p} \approx 1$.
The above solution was used as the initial condition
in the Crank-Nicholson finite difference code
(e.g., Press et al.\  1992)
that included the effects of dissipation.
All runs of the code reported in this paper utilized an
evenly spaced grid in $\ln k_{\perp}$, spanning six
orders of magnitude from $k_{\perp} R_{p} = 10^{-3}$ to
$10^{3}$.
The portions of the wave spectrum below the minimum
value of $k_{\perp}$ were assumed to remain at the
values given by the numerical integration of
eq.~(\ref{eq:Qquad}).
The implicit nature of the finite-difference scheme
allowed us to use a natural time step of 0.1 times the
minimum value of $\tau_s$ in the wavenumber grid.
The system was evolved toward a steady state, and any
numerical diffusivity was identified by simultaneously
evolving the same spectrum both with ($D \neq 0$) and
without ($D=0$) damping.
The latter case isolated long-time-scale numerical
effects that were negligible in magnitude, but nevertheless
divided out of the final damped spectrum.

An interesting feature of eqs.~(\ref{eq:dWdt}) and
(\ref{eq:dQdt}) is that the behavior of the inertial range
is largely independent of the values of $\beta$ and $\gamma$.
Indeed, even in the limit of pure advection (i.e.,
$\gamma = 0$), one can derive analytic solutions for the
inertial range and the dissipation range.
(Some of the supplementary quantities defined in
this Appendix become undefined for $\gamma = 0$, but the
original advection-diffusion equation remains well-defined.)
For $\gamma = 0$, the rapid cutoff in the dissipation range
is an explicit exponential decline (see, e.g., Townsend 1951;
Pao 1965).

The remainder of this Appendix is devoted to the derivation
of eq.~(\ref{eq:kapfun}), the analytic solution for $g(y)$.
Beginning with the time-steady advection-diffusion equation
(eq.~[\ref{eq:dEdt}]) in the inertial range ($S=D=0$), we
substitute in the {\em ansatz} functional form for
$W(k_{\perp}, y)$ given in eq.~(\ref{eq:Wgdef}).
Then, to evaluate the partial derivatives with respect to
$k_{\perp}$, we assume that $W_{\perp}$ conforms to a
local power-law form proportional to $k_{\perp}^{-2\zeta}$
(as defined in eq.~[\ref{eq:zetadef}]).
Even if $\zeta$ is allowed to vary as a ``slow'' function
of $k_{\perp}$, the dominant variation in $W_{\perp}$ should
remain the power-law decline with the local value of $\zeta$.
This assumption allows $\zeta$ to be considered constant
as a function of $y$.
After considerable but straightforward algebra,
eq.~(\ref{eq:dEdt}) becomes the following ordinary differential
equation with $y$ as the independent variable:
\begin{equation}
  (2 \zeta - \eta) \, h(y) \, + \, (1 - \zeta) \, y \, h'(y)
  \, + \, \alpha \, g''(y) \, = \, 0  \,\,\, .
  \label{eq:gODE}
\end{equation}
The supplementary quantity $h$ is defined as
\begin{equation}
  h(y) \, \equiv \, \left[ \beta + \gamma (1 + \zeta) \right]
  g(y) \, + \, \gamma (1 - \zeta) \, y \, g'(y)
  \,\,\, ,
\end{equation}
and primes denote differentiation with respect to $y$.
The dimensionless exponent $\eta$ is defined as
\begin{equation}
  \eta \, \equiv \, \frac{k_{\perp}}{2 \, \phi_e}
  \, \frac{\partial \phi_e}{\partial k_{\perp}}
\end{equation}
(i.e., $\eta = 0$ in the MHD inertial range and
$\eta = 1$ in the KAW inertial range).
Serendipitously, the ratio
\begin{displaymath}
  \frac{2\zeta - \eta}{1 - \zeta}
\end{displaymath}
is equal to 1 in both the MHD and KAW inertial ranges.
From numerical solutions of $W_{\perp}$ and $\phi_e$, the
above ratio is never less than 1 or larger than $\sim$1.3 in
the narrow transition between the MHD and KAW inertial ranges.
Thus, it seems reasonably safe to assume that this ratio is
unity for all relevant wavenumbers, and we simplify
eq.~(\ref{eq:gODE}) as
\begin{equation}
  ( 1 - \zeta ) \left[ h(y) \, + \, y \, h'(y) \right] \, + \,
  \alpha \, g''(y) \, = \, 0  \,\,\, .
\end{equation}
This equation is integrated once, assuming $g'(0) = 0$,
to obtain the first-order differential equation
\begin{equation}
  ( 1 - \zeta ) \, y \, h(y) \, + \, \alpha \, g'(y) \, = \, 0
  \,\,\, .
\end{equation}
Inserting the definition of $h(y)$, we obtain the relatively
simple equation
\begin{equation}
  \frac{1}{g} \frac{dg}{dy} \, = \,
  \frac{- (1 - \zeta) \, [ \beta + \gamma (1 + \zeta)] \, y}
  {\alpha + \gamma (1-\zeta)^{2} \, y^2}
\end{equation}
which can be integrated analytically to obtain
eq.~(\ref{eq:kapfun}).
For the aforementioned case of pure perpendicular advection
(i.e., $\gamma = 0$), $g(y)$ is a Gaussian function.
A larger value of $\gamma$ implies a stronger power-law tail
for $y \gg 1$.

If we had allowed the boundary condition $g'(0)$ to be a
nonzero constant, the resulting solutions for $g(y)$ would
have been asymmetric about $k_{\para} = 0$.
Na\"{\i}vely, these solutions could be considered to be
applicable to the relevant case of more power in outward
propagating fluctuations than in inward propagating
fluctuations.
However, the solutions exhibit unphysically negative
values of $g$ throughout most of the ``weaker'' half of
the distribution, and thus are probably inapplicable.

\vspace*{0.10in}
\footnotesize
\baselineskip=11pt
\setlength{\columnsep}{0.67cm}
\begin{multicols}{2}

\end{multicols}

\clearpage

\begin{figure}
\epsscale{0.63}
\plotone{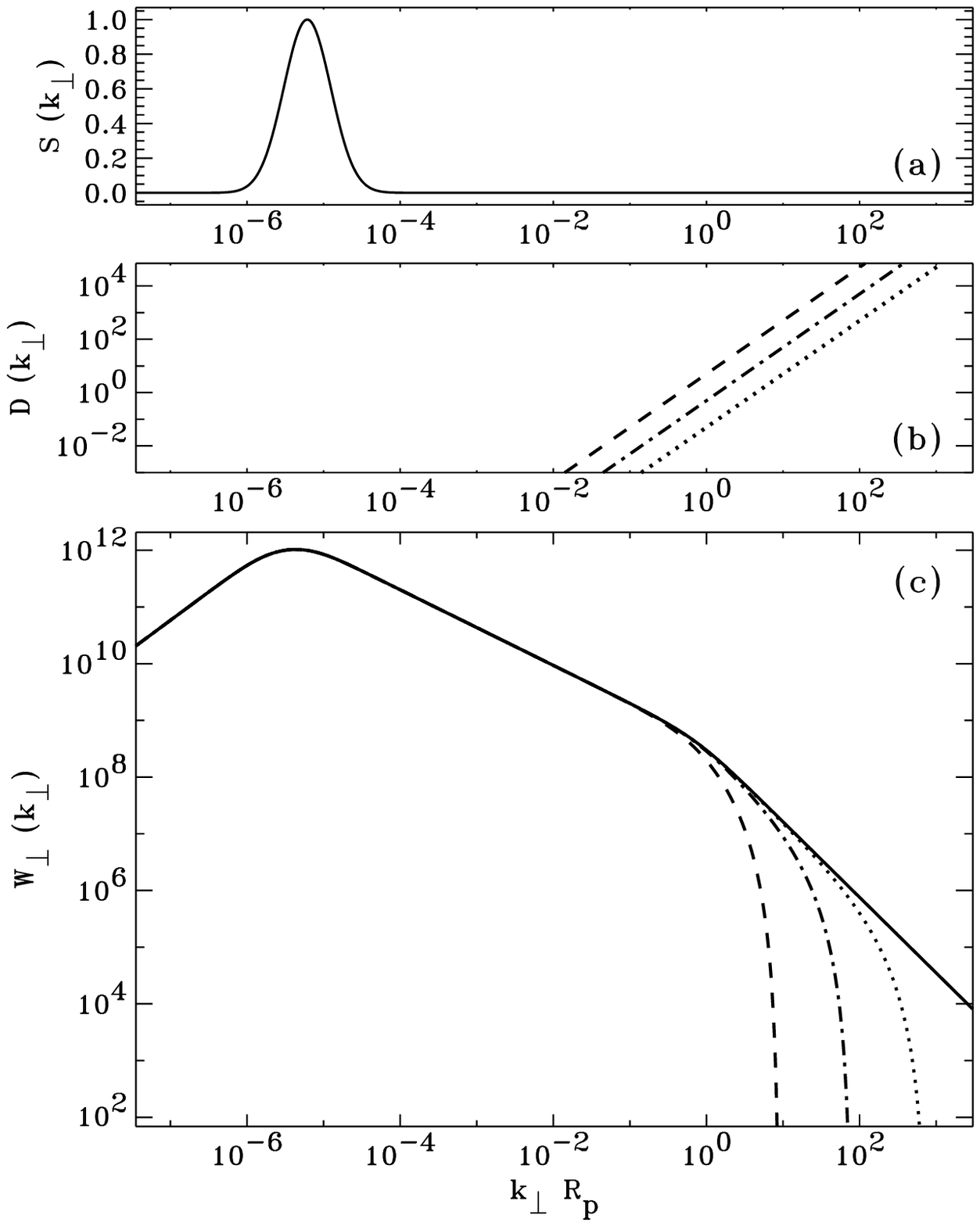}
\caption{\small
Representative solutions for the dominant
perpendicular cascade at $r = 2 \, R_{\odot}$: (a) outer-scale
energy source function $S(k_{\perp})$, normalized to its
maximum value; (b) dissipation rate
$D(k_{\perp}) = D_{0} ( k_{\perp} R_{p} )^{2}$, where
$D_{0} = 0.05$ s$^{-1}$ ({\em dotted line}),
0.5 s$^{-1}$ ({\em dot-dashed line}), and
5 s$^{-1}$ ({\em dashed line}); (c) time-steady reduced
power spectrum $W_{\perp} (k_{\perp})$ for the three
dissipation rates above, and for zero dissipation
({\em solid line}).  The proton gyroradius $R_p$ at this
radius in the extended corona is assumed to be equal to
0.024 km.}
\end{figure}

\clearpage

\begin{figure}
\epsscale{0.6}
\plotone{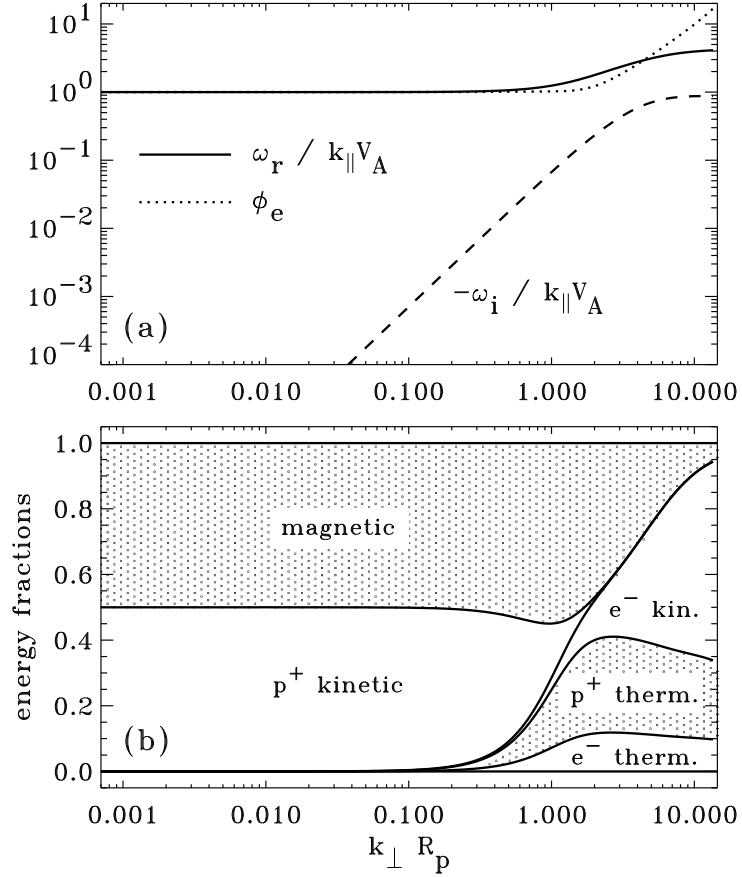}
\caption{\small
Dispersion and energy density properties of kinetic
Alfv\'{e}n waves as a function of $k_{\perp}$:
(a) real ({\em solid line}) and imaginary ({\em dashed
line}) parts of the wave frequency, scaled by the
MHD Alfv\'{e}n frequency, and the dimensionless
energy parameter $\phi_e$ ({\em dotted line});
(b) linearized energy density fractions---expressed as
areas---from top to bottom: magnetic field,
proton kinetic (i.e., velocity perturbation),
electron kinetic, proton thermal (i.e., density
perturbation), and electron thermal.
All curves correspond to a constant
value of $k_{\para} V_{A} / \Omega_{p} = 10^{-3}$.}
\end{figure}

\clearpage

\begin{figure}
\epsscale{0.6}
\plotone{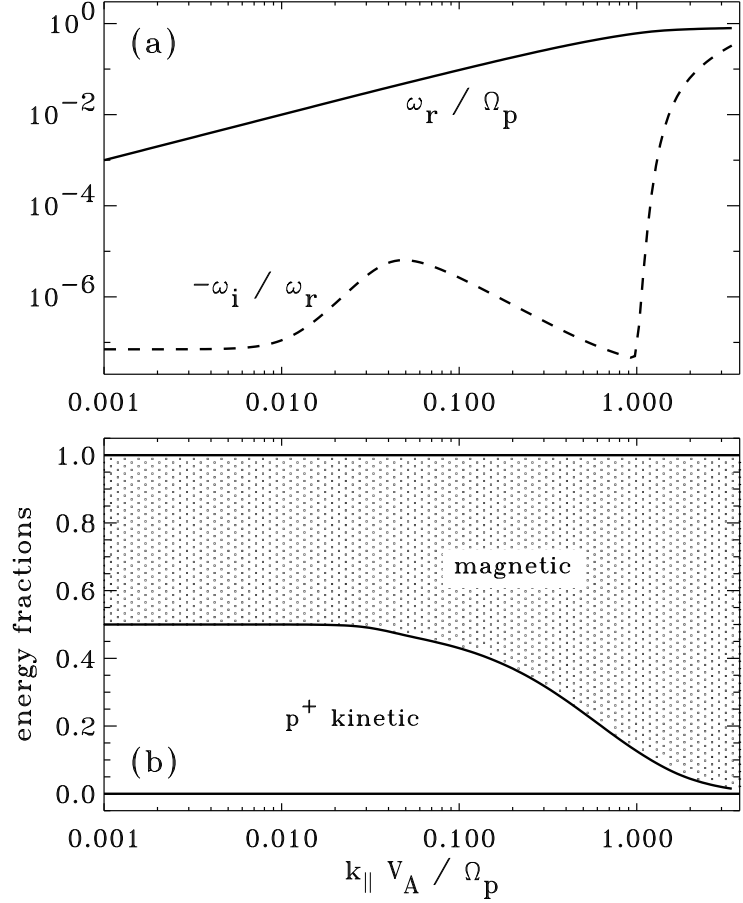}
\caption{\small
Dispersion and energy density properties of nearly
parallel-propagating Alfv\'{e}n waves (approaching
ion cyclotron resonance) as a function of $k_{\para}$.
Curves in (a) and areas in (b) are similar to those
in Figure 2, but the normalizations in (a) are
different ({\em see plot annotations}).
All curves correspond to a constant value of
$k_{\perp} R_{p} = 10^{-3}$.}
\end{figure}

\clearpage

\begin{figure}
\epsscale{0.85}
\plotone{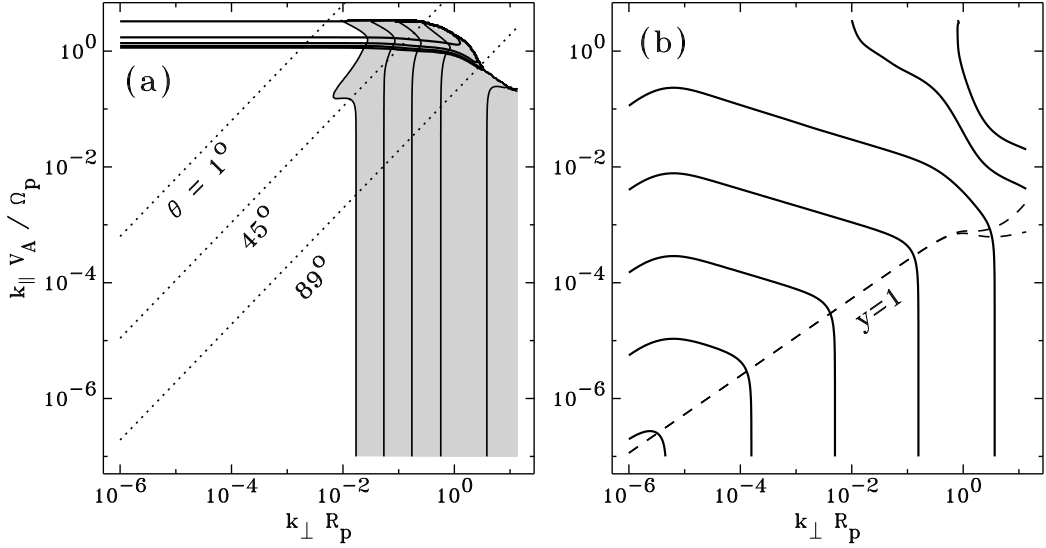}
\caption{\small
(a) Dimensionless damping rates
$| \omega_{i,s} / \omega_{r} |$
for protons ({\em unfilled contours}) and
electrons ({\em filled contours}) plotted one per decade
between $9 \times 10^{-5}$ and $9 \times 10^{-1}$ times the
maximum values of $| \omega_{i,s} / \omega_{r} |$ of 0.329
(protons) and 0.231 (electrons).
Both sets of contours go from low to high values with
increasing wavenumber.
Dotted lines show contours in $\theta$, the angle between
${\bf B}_0$ and ${\bf k}$.
(b) Contours of $W({\bf k})$, plotted one per $10^5$
between $10^{10}$ and $10^{40}$ cm$^5$ s$^{-2}$
({\em solid lines}) with the largest values at lower left.
Critical balance curves also plotted for both the undamped
({\em upper dashed line}) and damped ({\em lower dashed
line}) calculations of $W_{\perp} (k_{\perp})$.}
\end{figure}
\begin{figure}
\epsscale{0.57}
\plotone{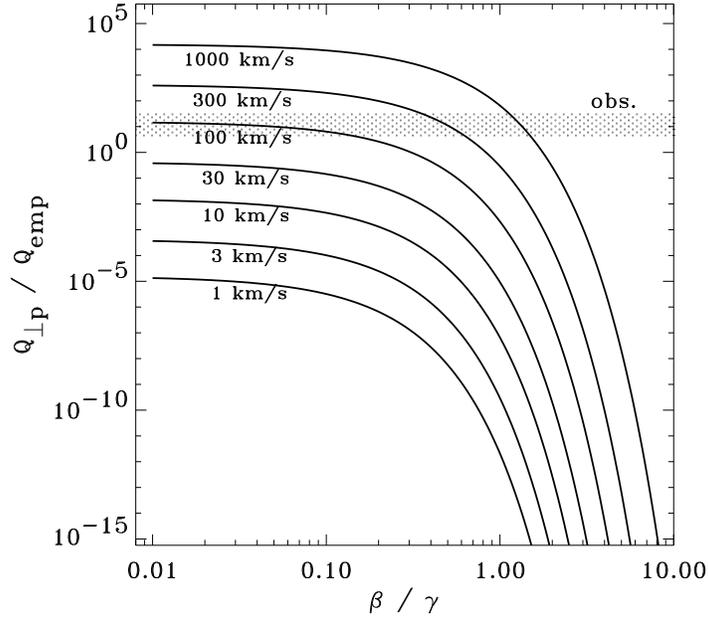}
\caption{\small
Perpendicular proton heating rates, in units of a fiducial
empirical rate $Q_{\rm emp}$ (see text), plotted versus
$\beta / \gamma$ for a range of total wave amplitudes
$(\delta U / \rho)^{1/2}$ (see labels on each curve).
The hashed region denotes empirical rates from
Li et al.\  (1999).}
\end{figure}

\clearpage

\begin{figure}
\epsscale{0.57}
\plotone{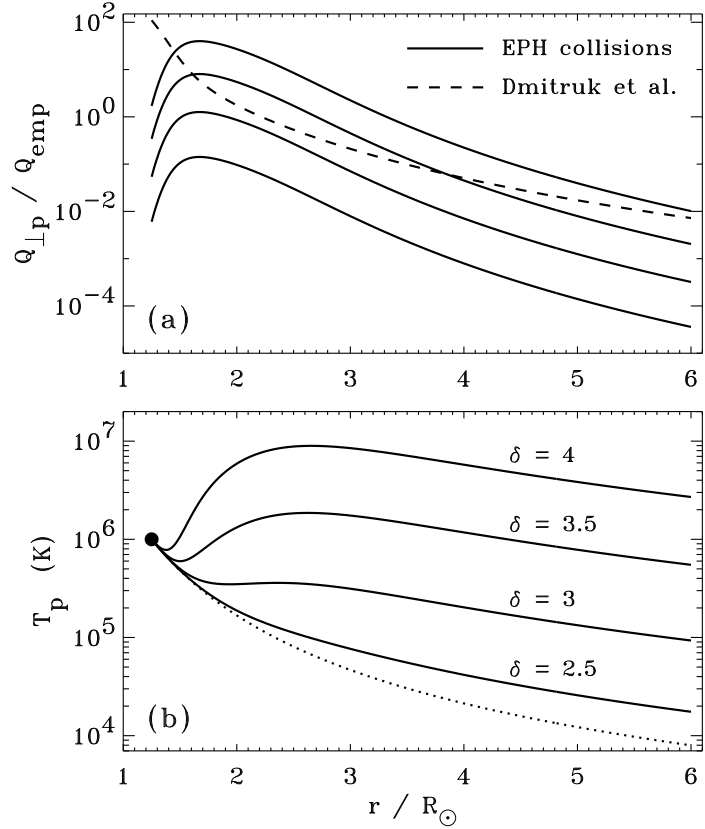}
\caption{\small
(a) Normalized proton heating rates plotted versus heliocentric
radius, both for the turbulent cascade rate of
Dmitruk et al.\  (2002) ({\em dashed line}) and for the derived
ion/EPH diffusion coefficient ({\em solid lines}).
From bottom to top, $\delta = 2.5$, 3, 3.5, 4.
(b) Perpendicular proton temperatures derived with the ion/EPH
heating rates plotted in (a) ({\em solid lines}), and
with no imposed heating (i.e., adiabatic cooling only;
{\em dotted line}).}
\end{figure}
\begin{figure}
\epsscale{0.9}
\plotone{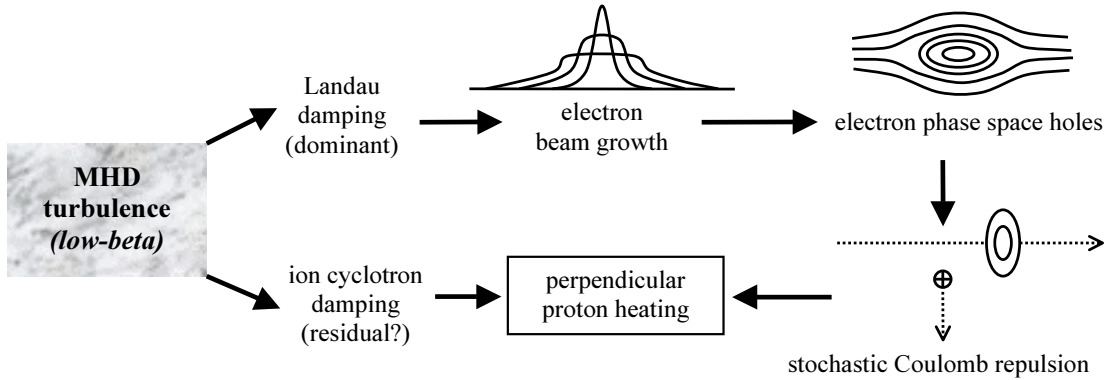}
\caption{\small
Schematic representation of the major physical processes
discussed in this paper.  The relative amount of turbulent
dissipation that directly heats protons and electrons
depends, among other factors, on the ratio $\beta / \gamma$.
The nonlinear development of parallel electron beams into
phase-space holes that can interact with protons is
also illustrated.}
\end{figure}

\end{document}